\documentclass[a4paper,11pt]{article}
\pdfoutput=1
\usepackage{jcappub} 
\usepackage[T1]{fontenc}
\usepackage{float}
\usepackage{hyperref}
\usepackage{comment}
\usepackage{tikz}
\usetikzlibrary{decorations.pathmorphing}

\usepackage{graphicx}
\usepackage{subcaption}
\usepackage{amsmath,amssymb}
\usepackage{multirow}
\usepackage{multicol}
\usepackage{mathtools}
\usepackage{orcidlink}

\usepackage[normalem]{ulem}

\subheader{\hfill \small CETUP-2025-019}

\title{Ruling Out Spiky WIMP Dark Matter using Indirect Searches}

\author[]{Dibya~S.~Chattopadhyay$^1$\orcidlink{0000-0003-2323-3950},}
\author[]{P.~S.~Bhupal~Dev$^2$\orcidlink{0000-0003-4655-2866},}
\author[]{Yago~Porto$^3$\orcidlink{0000-0003-3278-0948}}

\affiliation[1]{Department of Physics, Oklahoma State University, Stillwater, OK 74078, USA}
\affiliation[2]{Department of Physics and McDonnell Center for the Space Sciences, Washington University, St. Louis, MO 63130, USA}
\affiliation[3]{Physik-Department, Technische Universität München,
    James-Franck-Straße 1, 85748 Garching, Germany}

\emailAdd{dibya.chattopadhyay@okstate.edu}
\emailAdd{bdev@wustl.edu}
\emailAdd{yago.porto@tum.de}

\abstract{
The dark matter (DM) density profile in the innermost region of the Galaxy remains an open question.
In particular, while adiabatic growth of the supermassive black hole Sgr A$^\ast$ at the Galactic Center (GC) can induce a `spike' in central DM density, the existence of such a spike is still under debate.
Here we present new constraints on the spike slope $\gamma_{\rm sp}$ using conventional DM indirect detection searches.
We first recast existing photon and neutrino line searches, which include the contribution from the GC region, into constraints on the thermally-averaged DM annihilation cross section $\langle\sigma v\rangle$ in the presence of a DM spike.
We then derive new bounds on the spike profile for a generic Weakly Interacting Massive Particle (WIMP) DM scenario, where the thermal freeze-out mechanism fixes the annihilation cross-section at $\langle\sigma v\rangle\sim (2-3) \times 10^{-26}~{\rm cm}^3~{\rm s}^{-1}$.
We find that, for DM annihilation to photons, existing \emph{Fermi}-LAT and MAGIC data place strong constraints on spike profiles at the GC over a broad range of WIMP DM masses, from 10 GeV to 100 TeV, for photon branching fractions down to $\sim 10^{-2}$, and remain sensitive to values as small as $\sim 10^{-4}$ in parts of the parameter space.
For the neutrino channel, we use the recent IceCube data to constrain the existence of an extremely steep spike in the $\mathcal{O}(1-10)$~TeV DM mass range. Our analysis can be easily extended to other annihilation channels.}

\begin{document}
\maketitle
\flushbottom
\section{Introduction}
The presence of non-baryonic dark matter (DM) is supported by a wide range of cosmological and astrophysical observations, yet its particle nature remains unknown~\cite{Cirelli:2024ssz}.
Theoretically, perhaps the best-motivated scenario is that the DM is a weakly interacting massive particle (WIMP) produced thermally in the early Universe~\cite{Kolb:1990vq}.
In the standard freeze-out picture, DM remains in thermal equilibrium with the primordial plasma until the annihilation rate drops below the Hubble expansion rate.
The late-time abundance is then set predominantly by the thermally-averaged annihilation cross section $\langle \sigma v\rangle$, leading to the canonical estimate for the ``thermal relic freeze-out'' value of $\langle\sigma v\rangle \approx 3\times 10^{-26}\,{\rm cm^3\,s^{-1}}$ for $s$-wave annihilation~\cite{Kolb:1990vq}. A more precise treatment of the thermal relic freeze-out scenario leads to a required value of $\langle\sigma v\rangle_{\rm th} \approx 2.2\times 10^{-26}\,{\rm cm^3\,s^{-1}}$ for DM masses above $10$~GeV~\cite{Steigman:2012nb}.

Thermal WIMPs are constrained by a broad program of direct~\cite{Billard:2021uyg} and collider~\cite{Boveia:2018yeb} searches, as well as cosmological and indirect probes~\cite{PerezdelosHeros:2020qyt} which directly constrain $\langle \sigma v \rangle$.
At high redshifts, CMB anisotropies constrain energy injection from DM annihilation, providing robust limits on $s$-wave annihilation into a variety of Standard Model (SM) final states~\cite{Slatyer:2015jla}.
At late times, gamma-ray observations place stringent constraints over a wide mass range~\cite{Cooley:2022ufh}; in fact, searches in dwarf spheroidal galaxies provide some of the most robust limits on hadronic and leptonic channels (e.g.\ $b\bar b$ and $\tau^+\tau^-$), reaching the thermal scale for ${\cal O}(10\text{--}100)\,{\rm GeV}$ WIMPs~\cite{Fermi-LAT:2015att, McDaniel:2023bju}. Meanwhile, observations of the inner regions of the Milky Way by imaging atmospheric Cherenkov telescopes probe the TeV mass range and can also reach the canonical thermal benchmark for several SM final states under cuspy halo assumptions~\cite{CTA:2020qlo,HESS:2022ygk,Ryan:2023yzu,Alfaro:2025fxy}.

For line-like signatures, DM annihilation into photons typically enables some of the most stringent constraints on $\langle\sigma v\rangle$. 
Dedicated searches in the Galactic Center (GC) region with \emph{Fermi}-LAT data~\cite{Foster:2022nva} covering $m_\chi\sim(10- 2\times 10^3)~{\rm GeV}$, and MAGIC data~\cite{MAGIC:2022acl} extending up to $m_\chi\sim 10^5~{\rm GeV}$, place stringent constraints on $\chi\chi \to \gamma\gamma$ annihilation rate, ruling out thermal WIMPs for $m_\chi\lesssim 10^4$ GeV, if $\gamma\gamma$ is the dominant annihilation channel.
In contrast, DM annihilation into neutrinos~\cite{Beacom:2006tt, Yuksel:2007ac, Blennow:2019fhy,Arguelles:2019ouk} is far more challenging to probe and therefore generally yields substantially weaker bounds. Nevertheless, neutrino telescopes provide a complementary SM window that allows one to probe the DM annihilation to neutrinos in the GC. IceCube has performed a dedicated search for neutrino lines from $\chi\chi\to\nu\bar\nu$, constraining the corresponding annihilation cross section over a wide range of $m_\chi \sim (10 - 5\times 10^4)~{\rm GeV}$~\cite{IceCube:2023ies}. Similar neutrino line searches have also been performed by Super-Kamiokande~\cite{Super-Kamiokande:2020sgt} and ANTARES~\cite{ANTARES:2019svn}.

The DM annihilation signals crucially depend on the DM density distribution in the Galaxy. The local DM density is now known to a good accuracy from direct line-of-sight acceleration measurements of millisecond pulsars~\cite{Donlon:2025gxa}, which agrees well with the value inferred from stellar kinematics~\cite{Lim:2023lss}. However, the DM density profile in the innermost region of the Galaxy, which is most relevant for the annihilation signal from the GC, is still an open question. It is rather challenging to determine the exact DM density distribution in the inner Galaxy either from observations~\cite{Sofue:2013kja,Lacroix:2018zmg, Shao:2018klg, Shen:2023kkm, Ou:2023adg, John:2023knt,GRAVITY:2024tth, Acevedo:2025rqu, Gustafson:2025ypo,Scarcella:2025dsh, Paul:2026ikn} or from numerical simulations~\cite{Lim:2023lss, Putney:2025mch}. 
Therefore, different DM density profiles, such as NFW~\cite{Navarro:1995iw, Navarro:1996gj,Navarro:2003ew}, gNFW~\cite{Fukushige:2000ar, Navarro:2003ew,Diemand:2008in,Ou:2023adg}, Hernquist~\cite{Hernquist:1990be},  Burkert~\cite{Burkert:1995yz}, Einasto~\cite{Einasto:1965czb, Retana-Montenegro:2012dbd} and isothermal~\cite{Read:2015sta, Robertson:2016xjh, Fischer:2023lvl, Yang:2022zkd, Tran:2024hry}, have been used in the literature to capture the possibility of cuspy or cored DM density distributions close to the GC.

The existence of a supermassive black hole (SMBH) at the GC is now well-established from precise measurements of the stellar orbits around Sgr A$^\ast$~\cite{Ghez:2008ms, Gillessen:2008qv}. Recently, the Event Horizon Telescope has confirmed this finding by direct imaging of the Sgr A$^\ast$ shadow~\cite{EventHorizonTelescope:2022wkp, EventHorizonTelescope:2022wok}. Due to the interplay between the SMBH and DM, the actual DM density profile at the GC may significantly differ from the DM-only halo profile.
Ref.~\cite{Gondolo:1999ef} showed that the adiabatic growth of the central SMBH can induce a steep enhancement of the surrounding DM profile and generate a significant inner overdensity, known as the ``Gondolo-Silk'' (GS) spike with a density profile $\rho_{\rm sp}(r)\propto r^{-\gamma_{\rm sp}}$, with the spike slope range $\gamma_{\rm sp}\in [2.25,2.5]$. The existence of such a DM spike in the vicinity of the GC can substantially boost the DM indirect detection signal, relative to expectations from a smooth halo, because the DM annihilation $J$-factor grows as $\rho_\chi^2$, where $\rho_\chi$ denotes the DM density.
The spike hypothesis has been widely explored in the indirect-detection literature for the Milky Way and for other galaxies hosting SMBHs, including early studies of annihilation radiation~\cite{Gondolo:2000pn,  Merritt:2002vj, Bertone:2002je,Bertone:2005hw}, synchrotron-based probes~\cite{Lacroix:2013qka}, and applications to gamma-ray~\cite{Regis:2008ij,Lacroix:2015lxa, Shelton:2015aqa, Sandick:2016zeg,Johnson:2019hsm, Chiang:2019zjj,Alvarez:2020fyo, Balaji:2023hmy, Christy:2023tdv,Yang:2024jtp,Chattopadhyay:2024qgs,Phoroutan-Mehr:2024cwd,Egorov:2025dey},  neutrino~\cite{Cline:2022qld, 
Ferrer:2022kei, Cline:2023tkp, 
Akita:2025dhg} and boosted DM~\cite{Wang:2021jic, Granelli:2022ysi, DeMarchi:2025uoo, Barillier:2025xct} signals coming from SMBH environments (like AGNs), as well as future supernova neutrino signals~\cite{Dev:2025tdv}. Further motivations for the existence of a DM spike come from the fact that it can potentially explain some anomalous indirect detection signals, such as the cosmic-ray electron excess~\cite{Sandick:2010yd, Coogan:2019uij, Chan:2025bxn}, GC gamma-ray excess~\cite{Hooper:2010mq,Fields:2014pia} or the Galactic bulge 511~keV emission profile~\cite{laTorreLuquePedro:2024est}, depending on the DM mass chosen.
Ref.~\cite{Chan:2024yht} claims evidence for a DM density spike around the SMBH binary OJ 287, where the best-fit spike slope is found to be $\gamma_{\rm sp}=2.351^{+0.032}_{-0.045}$, consistent with the GS adiabatic growth model.

However, the presence of a DM spike at \emph{our own GC}~has not been conclusively established yet~\cite{Ullio:2001fb, Lacroix:2018zmg,Shen:2023kkm, GRAVITY:2024tth, Meighen-Berger:2025hrq}. 
Moreover, even if a spike forms, it can be weakened by subsequent dynamical processes such as stellar heating, dynamical friction, and gravitational interactions with a companion to the SMBH~\cite{Merritt:2003qk, Lopez:2025eqz}. Strong DM self-interactions, including number-changing interactions, can also substantially deplete and reshape DM spikes~\cite{BetancourtKamenetskaia:2025ivl}.
General-relativistic corrections to DM spike profiles and more realistic SMBH formation scenarios have been discussed in   Refs.~\cite{Sadeghian:2013laa, Caiozzo:2025mye}. 
Effectively, the question of whether a GC spike exists or not remains unresolved. Answering this important question can provide valuable clues about the Milky Way's dynamical history and the formation and growth of its central SMBH. Future gravitational-wave observations of compact binaries on accelerated orbits around SMBHs have been proposed as a clean probe of DM density spikes, allowing measurements of the spike slope~\cite{Feng:2025fkc,Li:2025qtb,Tiwari:2025qqx}. It was also recently shown~\cite{Meighen-Berger:2025hrq} that DM spikes around SMBHs can affect the survival and composition of ultra-high-energy cosmic rays, providing a novel way to probe DM-nucleon interactions.

In this work, we quantify how a GC spike would modify DM annihilation signals. Although naively this might seem like a trivial rescaling of numbers, we show that it is more complicated because the spike profile, and hence the $J$-factor, depends on the DM mass and annihilation rate.
After carefully taking this into account, we use the existing data from DM indirect detection searches and compare it with the thermal WIMP freeze-out expectation to derive new constraints on the spike slope parameter   $\gamma_{\rm sp}$ as a function of the WIMP DM mass.
We use the recent GC line searches in $\gamma$ rays and neutrinos, and recast the $\langle \sigma v\rangle$ constraints obtained under the assumption of a smooth NFW halo profile~\cite{Navarro:1995iw} into constraints for spiky DM profiles~\cite{Gondolo:1999ef}.
Specifically, we recast the GC $\chi\chi \to \gamma\gamma$ limits from \emph{Fermi}-LAT~\cite{Foster:2022nva} and MAGIC~\cite{MAGIC:2022acl}, as well as the IceCube GC $\chi\chi \to \nu \bar \nu$ limits~\cite{IceCube:2023ies}, to derive the corresponding constraints in spiky-halo scenarios. We limit ourselves to these experiments because they did not mask the GC (unlike, e.g.~Refs.~\cite{HESS:2022ygk, Alfaro:2025fxy}).
Under the popular assumption of a purely thermal-relic freeze-out WIMP DM scenario, we then translate our rescaled constraints on $\langle\sigma v\rangle$ into stringent constraints on the GS spike parameter $\gamma_{\rm sp}$ and show that the GS spike with $\gamma_{\rm sp}\in [2.25,\,2.5]$ is ruled out for a wide range of DM mass $m_\chi\in [10,\,10^5]$ GeV, provided $\chi\chi\to \gamma\gamma$ contributes $\gtrsim 1\%$ of the total DM annihilation rate. The corresponding constraints derived from the neutrino channel are relatively weaker, as expected, but still provide a complementary probe of $\gamma_{\rm sp}$ in a limited range of $m_\chi\in [1800,\,7900]$~GeV.

The rest of the paper is organized as follows: in section~\ref{sec:spike-profile}, we discuss the shape of the DM profile considered, and calculate how the DM annihilation $J$-factor behaves as a function of DM mass, cross section, and the spike slope.
In section~\ref{sec:gamma_lines} we present the results for DM annihilation into photons, and put constraints on the spike parameter $\gamma_{\rm sp}$.
In section~\ref{sec:nu_lines} we repeat the same procedure for DM annihilation into neutrinos.
We conclude in section~\ref{sec:conclusion}.
In Appendix~\ref{sec:appendixA}, we give some details of the $J$-factor dependence on the DM parameters.

\section{DM spike around Sgr~A$^\ast$}
\label{sec:spike-profile}

In this section, we discuss how a GS spike modifies the DM distribution around Sgr A$^\ast$ -- the central SMBH of our Milky Way galaxy. We then discuss the effects of the spike on the DM annihilation $J$-factor.

\subsection{DM density profile}
\label{subsec:dm_density}

A DM density spike may form via the adiabatic growth of the SMBH at the center of a pre-existing halo cusp~\cite{Gondolo:1999ef}.
In this scenario, DM orbits contract as the central potential deepens, generating an inner power-law enhancement over a characteristic radius $R_{\rm sp}$.
Denoting the pre-spike inner cusp as $\rho_{\rm halo}(r)\propto r^{-\gamma}$ (with $\gamma=1$ for NFW, and $\gamma=0$ for a more cored profile), adiabatic growth of the SMBH and the associated adiabatic contraction of collisionless DM orbits yield an inner spike that is well approximated by a power law:
\begin{equation}
\rho_{\rm sp}(r)=\rho_{\rm halo}(R_{\rm sp})\left(\frac{r}{R_{\rm sp}}\right)^{-\gamma_{\rm sp}},
\qquad  r \lesssim R_{\rm sp},
\label{eq:spike_powerlaw}
\end{equation}
with a slope $\gamma_{\rm sp}$ given by $\gamma_{\rm sp}=(9-2\gamma)/(4-\gamma)$ for the GS case, $R_{\rm sp}$ denotes the size of the spike. For a range of cored initial profiles corresponding to $\gamma = 0$ to a much more cuspy initial condition of $\gamma=2$, we obtain a final value of $\gamma_{\rm sp}$ in the range of $[2.25,\,2.5]$.

Note that the power law behavior of the inner enhancement cannot persist down to arbitrarily small radii toward the SMBH. Sufficiently close to the center, additional physical effects may regulate the spike and truncate the simple power-law behavior. 
First, DM self-annihilations can deplete the inner density and generate an approximately constant ``saturation’’ (or annihilation) plateau.\footnote{This is not applicable if the DM is non-annihilating (e.g., asymmetric~\cite{Zurek:2013wia}), where effectively we have $\langle \sigma v\rangle\to 0$. But in this case, we lose the indirect detection signal we are relying on here to probe the spike.} A useful estimate follows by requiring that the local annihilation timescale $t_{\rm ann}$ not be shorter than the spike age $t_\ast$. Since the per-particle annihilation rate scales as $\Gamma_{\rm ann}(r)\sim n(r)\langle\sigma v\rangle$ with $n=\rho_{\rm sp}/m_\chi$, the corresponding timescale is
\begin{equation}
t_{\rm ann}(r) \equiv \Gamma_{\rm ann}^{-1}(r)
\sim \frac{m_\chi}{\rho_{\rm sp}(r)\langle\sigma v\rangle}\,.
\end{equation}
Imposing $t_{\rm ann}\gtrsim t_\ast$ then implies a maximal (saturation) density,
\begin{equation}
\rho_{\rm sat} \approx \frac{m_\chi}{\langle\sigma v\rangle t_\ast},
\label{eq:saturation_density}
\end{equation}
so that the inner profile is effectively capped at $\rho_{\rm sat}$ and the transition radius $R_{\rm sat}$ is defined implicitly by $\rho_{\rm sp}(R_{\rm sat})=\rho_{\rm sat}$. For Milky Way, we take $t_\ast = 10^{10}$~years, which leads to an estimate for the saturation density as
\begin{equation}
    \rho_{\rm sat} \approx 1.44 \times 10^{9}~{\rm GeV\,cm}^{-3} \bigg( \frac{m_{\chi}}{10~{\rm GeV}}\bigg) \bigg( \frac{2.2 \times 10^{-26}~{\rm cm}^3~{\rm s}^{-1}}{\langle \sigma v \rangle}\bigg)\;.
\end{equation}
Second, the spike may be further softened by gravitational interactions with the dense stellar environment near the GC and by the SMBH merger history, both of which can partially erase a steep cusp \cite{Ullio:2001fb, Merritt:2003qk}. Modern treatments, therefore, consider families of spike profiles with varying $\gamma_{\rm sp}$ and $R_{\rm sp}$, including scenarios where the central overdensity is substantially reduced. In our analysis, we will treat the spike slope $\gamma_{\rm sp}$ as a free parameter, allowing it to vary between $1$ (corresponding to the standard NFW scenario) and $2.5$ (corresponding to the maximum GS spike value).

For definiteness, we adopt an explicit Milky Way halo+spike profile closely following the parametrization in Ref.~\cite{laTorreLuquePedro:2024est}:\footnote{A more accurate treatment of the adiabatic growth of the DM spike, including general relativistic effects, can extend the spike down to $2R_S$~\cite{Sadeghian:2013laa}, as used in Refs.~\cite{Balaji:2023hmy, laTorreLuquePedro:2024est}. This enhancement is even more pronounced for a rotating Kerr black hole~\cite{Ferrer:2017xwm}. However, to be on the conservative side, we assume a Schwarzschild black hole with the spike density given by Eq.~\eqref{eq:MW_profile_piecewise} where the cut-off scale is chosen to be $4R_S$, as used e.g.~in Refs.~\cite{Gondolo:1999ef, Akita:2025dhg, Meighen-Berger:2025hrq}.}
\begin{equation}
\rho_\chi(r)= \left(1- \frac{4R_S}{r}\right)^3 \times
\begin{cases}
0\,, & r < 4R_S\,,\\[1pt]
\rho_{\rm sat}\left(\dfrac{r}{R_{\rm sat}}\right)^{-1/2}, & 4R_S \le r < R_{\rm sat}\,,\\[8pt]
\rho_{\rm sp}(r), & R_{\rm sat} \le r < R_{\rm sp}\,,\\[8pt]
\rho_{\rm halo}(r)\,, & r \ge R_{\rm sp}\,,
\end{cases}
\label{eq:MW_profile_piecewise}
\end{equation}
where $R_S\equiv 2GM_{\rm BH}/c^2$ is the Schwarzschild radius of the central SMBH and $G$ is the Gravitational constant. For Milky Way, the mass of Sgr~A$^\ast$ is $M_{\rm BH} = 4.3 \times 10^6 M_\odot$~\cite{GRAVITY:2021xju}, which gives a Schwarzschild radius of $R_S \simeq 1.26 \times 10^7$~km $\simeq 4.1\times 10^{-7}$ pc.  The spike radius $R_{\rm sp}$ is given by $R_{\rm sp} = 0.2 \,G M_{\rm BH}/ v_0^2$, where $v_0$ is the bulge velocity dispersion; for Milky Way, $v_0 = 105 \pm 20~{\rm km~s}^{-1}$~\cite{Gultekin:2009qn} leads to a spike radius of $R_{\rm sp} \approx 0.33$~pc.
For the outer halo, we use the NFW profile~\cite{Navarro:1995iw}:
\begin{equation}
\rho_{\rm halo}(r)\equiv\rho_{\rm NFW}(r)=\rho_s\left(\frac{r}{r_s}\right)^{-1}\left(1+\frac{r}{r_s}\right)^{-2}.
\label{eq:NFW_halo}
\end{equation}
For the NFW halo parameters, we employ the values used in Ref.~\cite{Foster:2022nva}: 
\begin{equation}
\label{eq:nfw_standard}
    r_s=20~{\rm kpc}\; , 
    \quad R_\odot = 8.5~{\rm kpc}\; , 
    \quad \rho_s \approx 0.345~{\rm GeV\, cm}^{-3} \; ,
\end{equation}
 with the density normalization fixed by a local DM density of $\rho_\odot=0.4~{\rm GeV\,cm^{-3}}$  at a distance  $R_\odot$ from the GC. We consider a DM halo size of $400$~kpc, although our results are not sensitive to this choice, since the DM density drops to negligible values beyond 20 kpc. In fact, the DM density at $400$~kpc is approximately $3.9\times 10^{-5}~{\rm GeV\,cm^{-3}}$; since the annihilation signal scales as $\rho_\chi^2$, the precise choice of the outer halo boundary has a negligible effect on our results.
 Note that different NFW parametrizations adopted in the literature require us to consistently recast the published constraints into our chosen NFW profile.

\begin{figure}[t]
    \centering
    \includegraphics[height=0.6\linewidth]{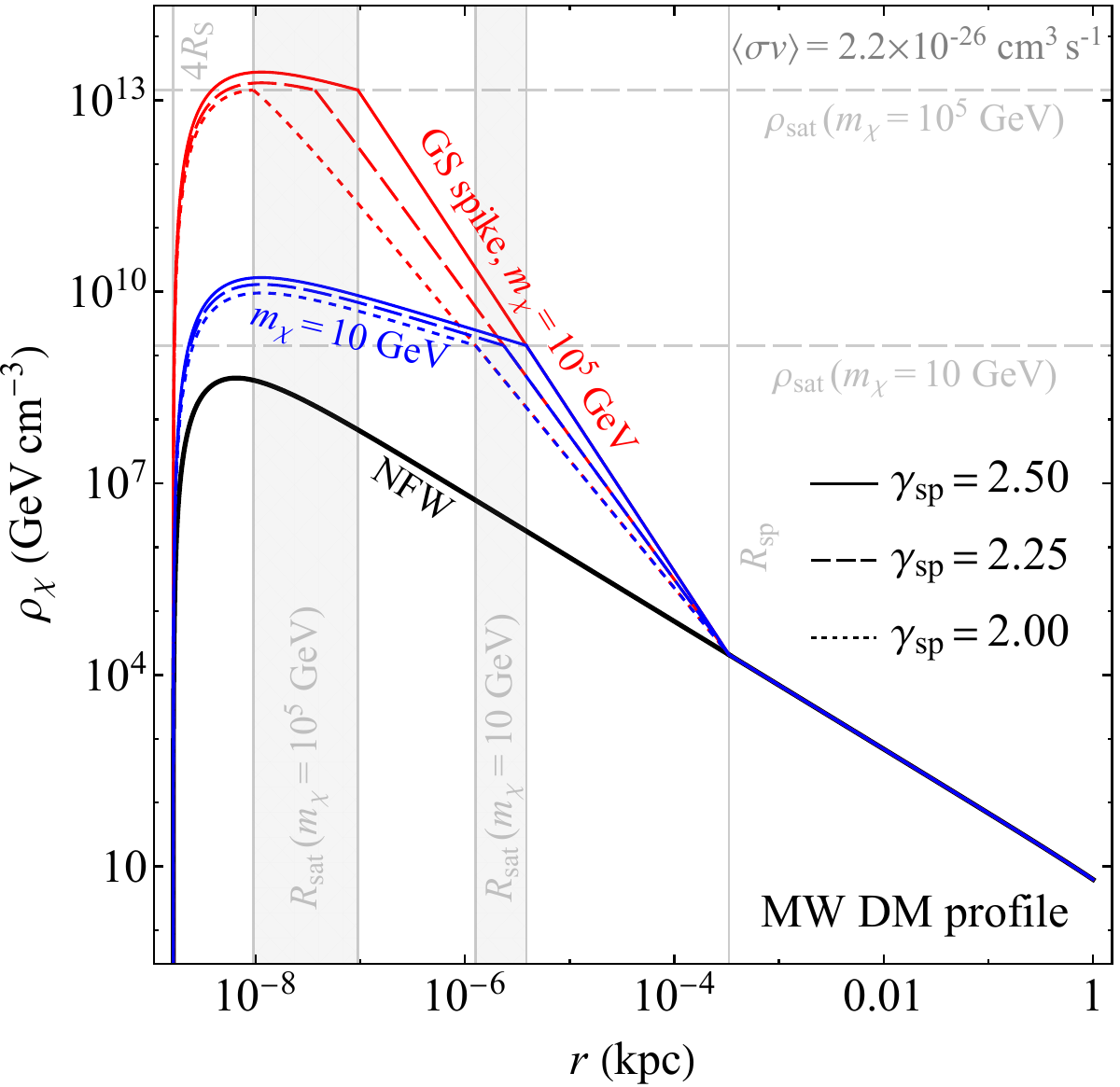}
    \caption{
    Illustration of the spike+halo construction for the Milky Way DM density as a function of the distance from the GC, for an NFW profile with DM mass of $m_\chi = 10$~GeV (black), and in the presence of DM spike (with slopes $\gamma_{\rm sp} =$ 2, 2.25 and 2.5, shown by the thin dotted, dashed and solid lines, respectively), for $m_\chi = 10$~GeV (blue), and $10^5$~GeV (red). The different spike parameters lead to different DM densities for $r \leq R_{\rm sp}$. The annihilation-regulated behavior for $r \leq R_{\rm sat}$ (the saturation radius) is illustrated for a fixed value of $\langle \sigma v\rangle = 2.2 \times 10^{-26}\;{\rm cm}^3\, {\rm s}^{-1}$. For larger DM masses, the value of the saturation density $\rho_{\rm sat}$ is higher, which leads to the DM spike continuing till smaller values of $R_{\rm sat}$.
    }
    \label{fig:dm_density}
\end{figure}

In Fig.~\ref{fig:dm_density}, we show the DM density profiles obtained for a standard NFW profile [cf.~Eq.~\eqref{eq:NFW_halo} multiplied by the relativistic correction factor $(1-4R_S/r)^3$], as well as for a GS spike [cf.~Eq.~\eqref{eq:MW_profile_piecewise}]  with representative values of $\gamma_{\rm sp} = 2, \, 2.25,$ and $2.5$.
Note that the density profiles for different DM masses coincide above the spike radius ($r \geq R_{\rm sp}$), and start diverging before reaching the saturation density, i.e., in the range $R_{\rm sat} \leq r \leq R_{\rm sp}$.
For larger masses, the spike profiles reach saturation density at smaller $R_{\rm sat}$ values, leading to an overall larger impact of the GS spike, with DM densities reaching values as high as $\rho_\chi \sim 10^{13}~{\rm GeV}~{\rm cm}^{-3}$ for $m_\chi = 10^5$~GeV. 
In Fig.~\ref{fig:dm_density}, we show the saturation densities, and the resulting range saturation radius ($R_{\rm sat}$) for different DM masses ($m_\chi = 10$~GeV, and $m_\chi = 10^5$~GeV), for spikes in the range of $\gamma_{\rm sp} \in [2,\,2.5]$. The downturn of the DM densities close to $4R_{S}$ is due to the relativistic suppression factor of $(1-4R_S/r)^3$ in Eq.~(\ref{eq:MW_profile_piecewise}).

For consistency, in our analysis, we also consider that the reference NFW curve can saturate at $\rho_{\rm sat}$ for low mass DM or large values of the annihilation cross section.
This differs from the common convention of using an unsaturated NFW profile in the literature (e.g.~\cite{Balaji:2023hmy, laTorreLuquePedro:2024est}).
However, for the benchmark masses considered here, this choice does not introduce any observable differences,  since, unless in the presence of a spiky DM profile, the contribution from the very inner regions of the GC is negligible. We also checked that using a slightly different version of the spike profile as in Refs.~\cite{Gondolo:1999ef, Akita:2025dhg, Meighen-Berger:2025hrq} does not lead to a significant change in our final results. 
On the other hand, using an unsaturated spiky profile (as e.g. in Ref.~\cite{Balaji:2023hmy}) would result in a much stronger constraint than those derived here. Even just ignoring the relativistic correction factor in Eq.~\eqref{eq:MW_profile_piecewise} gives us roughly two orders of magnitude stronger constraint on  $\langle\sigma v\rangle$.

\subsection{DM annihilation $J$-factor}
\label{subsec:Jfactor}
For DM annihilation into SM final states, the differential flux from a region of interest (ROI) is
\begin{equation}
\frac{d\Phi_{\rm SM}}{dE}(\Delta\Omega)
=
\frac{\langle\sigma v\rangle}{8\pi\, m_\chi^2}\,
\frac{dN_{\rm SM}}{dE}\,
J(\Delta\Omega),
\label{eq:flux_master}
\end{equation}
where $dN_{\rm SM}/dE$ is the SM final state yield per annihilation and $\Delta \Omega$ denotes the ROI solid angle.
All astrophysical dependence is contained within the line-of-sight (l.o.s.) annihilation $J$-factor, defined as
\begin{equation}
J(\Delta\Omega)
=
\int_{\Delta\Omega} d\Omega
\int_{\rm l.o.s.} ds\;
\rho_\chi^2\!\big[r(s,\theta)\big],
\label{eq:J_def}
\end{equation}
where the distance from the core $r(s,\theta)$ is given by
\begin{equation}
    r(s,\theta)=\sqrt{r_\odot^2+s^2-2r_\odot s\cos\theta}.
\end{equation}
Here, $\theta$ is the angle to the GC and $r_\odot$ the distance of Earth/Sun from the GC, $r_\odot=8.5$~kpc.
For the circular apertures used throughout this work, we parametrize the ROI by its half-opening angle $\theta_{\max}$, such that
\begin{equation}
\Delta\Omega \equiv \Delta\Omega(\theta_{\max}) = 2\pi\bigl(1-\cos\theta_{\max}\bigr)\; .
\end{equation}
In the limit of $\theta_{\max}\ll 1$, this expression simplifies to $\Delta\Omega \approx \pi\,\theta_{\max}^2$.

Experimental constraints derived under a reference $\rho_{\rm NFW}(r)$ effectively constrain the product $\langle\sigma v\rangle_{\rm NFW} \times J_{\rm NFW}$ for a given ROI $\Delta\Omega$.
In the presence of a GS spike, the saturation region (regulated by efficient annihilation of the DM) implies that $J_\chi(\Delta\Omega)$ depends not only on the slope of the spike $\gamma_{\rm sp}$, but also on the $\langle\sigma v\rangle$ through $\rho_{\rm sat}$ and $R_{\rm sat}$. In particular, a smaller (larger) value of $\langle\sigma v\rangle$ implies a larger (smaller) saturation density $\rho_{\rm sat}$ [cf.~Eq.~\eqref{eq:saturation_density}], which in turn shifts the saturation radius to smaller (larger) values. As a result, the spike contribution to the $J$-factor is enhanced (suppressed) for smaller (larger) $\langle\sigma v\rangle$. Further, since $\rho_{\rm sat} \propto m_{\chi}$, the mass of the DM also regulates the $J$-factor in the presence of a GS spike. 
We therefore recast the bound $\langle\sigma v\rangle_{\rm NFW}$ in presence of the spike by solving
\begin{equation}
\langle\sigma v\rangle_{\rm spike} \times 
J_{\rm spike}\left(m_\chi,\langle\sigma v\rangle_{\rm spike}, \Delta\Omega, \gamma_{\rm sp}\right)
=
\langle\sigma v\rangle_{\rm NFW} \times 
J_{\rm NFW}(\Delta\Omega),
\label{eq:recast_equation}
\end{equation}
where $J_{\rm spike}\left(m_\chi,\langle\sigma v\rangle_{\rm spike}, \Delta\Omega, \gamma_{\rm sp}\right)$ denotes the annihilation $J$-factor in the presence of a spike, and is computed from Eq.~\eqref{eq:J_def} using the halo+spike profile in Eq.~\eqref{eq:MW_profile_piecewise}.

We decompose the $J$-factor into a standard NFW halo contribution and a GS spike-induced factor, as
\begin{equation}
J_\chi(\Delta\Omega)=J_{\rm NFW}(\Delta\Omega)+\Delta J(\Delta\Omega),
\label{eq:Jdecomp_def}
\end{equation}
where,
\begin{equation}
    \Delta J(\Delta\Omega)\equiv \int_{\Delta\Omega} d\Omega \int_{\rm l.o.s.} ds\;
\left[\rho_\chi^2(r)-\rho_{\rm NFW}^2(r)\right].
\end{equation}
Since $\rho_\chi(r)=\rho_{\rm NFW}(r)$ for $r\ge R_{\rm sp}$ in Eq.~\eqref{eq:MW_profile_piecewise}, $\Delta J(\Delta\Omega)$ only has contribution from the region $r < R_{\rm sp}$ (as long as the ROI, regulated by $\Delta\Omega$, covers the full spiked region). In this inner region, the spiked component would typically dominate over the NFW contribution, except for values close to  $\gamma_{\rm sp} =1$. Therefore, the GS-spike contribution can be approximated to be
\begin{align}
    \Delta J \approx & \,\frac{1}{R_\odot^2} \int_{4 R_S}^{R_{\rm sp}} dr\, 4\pi r^2 \, \rho_{\rm sp}^2(r)
    \; , \nonumber\\
    \approx & \, \frac{4\pi}{R_\odot^2} \rho_{\rm halo}^2 (R_{\rm sp})  \left[  \frac{2\gamma_{\mathrm{sp}} - 1}{2\left(2\gamma_{\mathrm{sp}} - 3\right)}\, R_{\mathrm{sp}}^{3}\left(\frac{R_{\mathrm{sat}}}{R_{\mathrm{sp}}}\right)^{3-2\gamma_{\mathrm{sp}}} \right]
    \label{eq:jchi_scaling}
\end{align}
for a spike parameter of $\gamma_{\rm sp}>3/2$, and in the limit $R_{S} \ll R_{\rm sat} \ll R_{\rm sp}$.
For $\gamma_{\rm sp}>3/2$, the spike is sufficiently steep that the annihilation integral is dominated by the innermost region. This is because, in this limit, terms proportional to $R_{\rm sat}^{\,3-2\gamma_{\rm sp}}$ are enhanced for $R_{\rm sat}\ll R_{\rm sp}$. As a result, we find the contribution associated with the spike's outer boundary at $R_{\rm sp}$ to be subleading, while the dominant dependence arises from the inner cutoff $R_{\rm sat}$ (set by annihilation saturation).
In this regime, the spike contribution scales as
$J_{\rm spike}\propto R_{\rm sat}^{\,3-2\gamma_{\rm sp}}$, so that a smaller (larger) $R_{\rm sat}$ yields a larger (smaller) $J$-factor. Note, however, for larger DM masses and smaller values of the annihilation cross-section, the $R_{S} \ll R_{\rm sat}$ limit will not hold, and in this regime, the effects of the relativistic corrections will change the simple power-law behavior discussed above.

Eq.~\eqref{eq:jchi_scaling} highlights that modest changes in $\gamma_{\rm sp}$ can lead to large variations in the $J$-factor and thus appreciably modify the inferred limits. In the remainder of this work, we compute the  $J$-factor for the spike+halo profiles and determine the corresponding constraints by solving Eq.~\eqref{eq:recast_equation} for each ROI and dataset. We then translate these into bounds on the spike slope $\gamma_{\rm sp}$ for the thermal-relic benchmark choice of $\langle \sigma v\rangle$ as a function of $m_\chi$~\cite{Steigman:2012nb}.

\begin{figure}[t]
    \centering
    \includegraphics[height=0.6\linewidth]{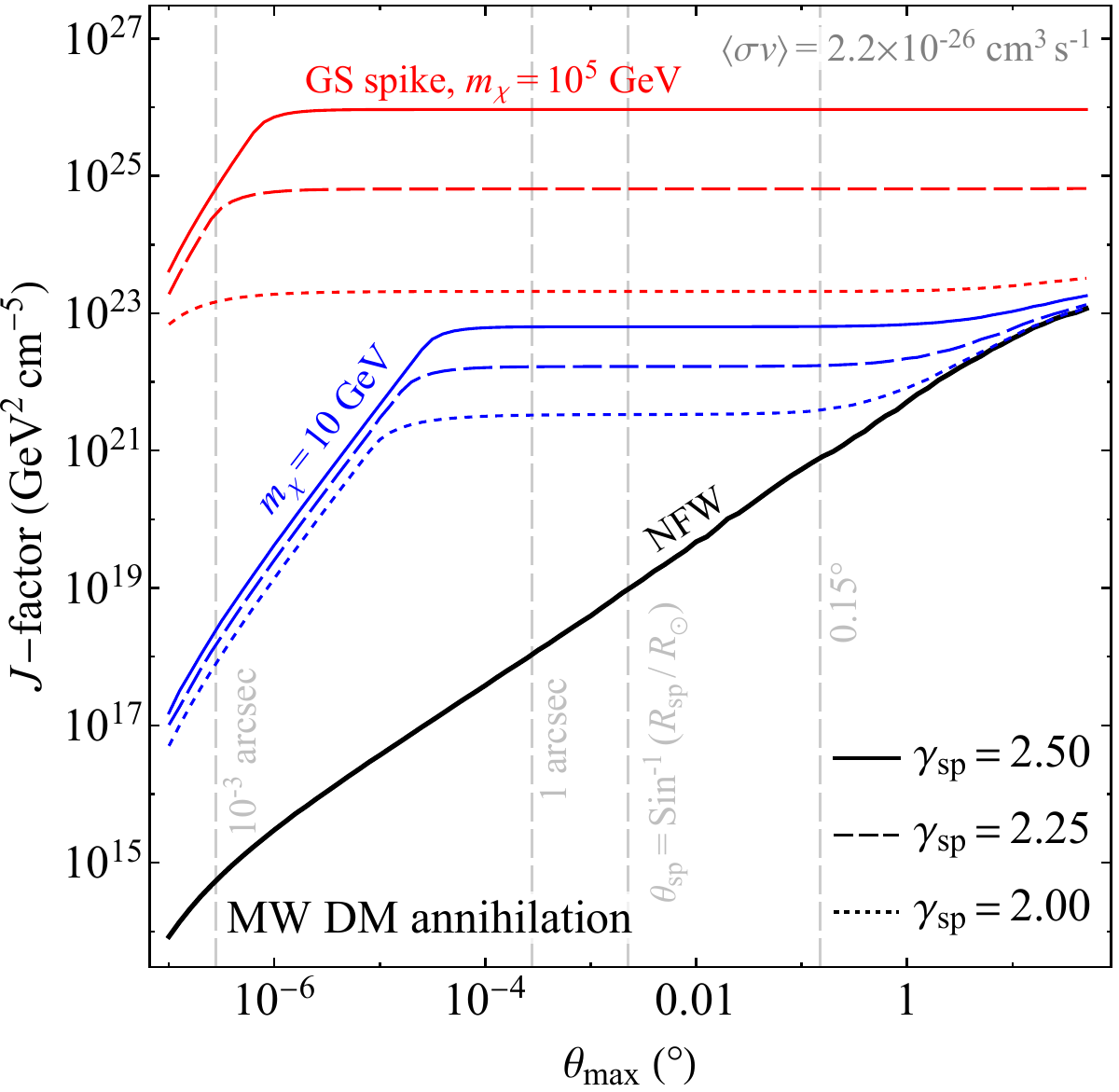}
    \caption{The impact of the spiky DM profile on the annihilation $J$-factor as a function of the maximum angular cutoff $\theta_{\max}$. We show results for an NFW profile with DM mass of $m_\chi \geq 10$~GeV (black curve), and in the presence of a DM spike (with slopes $\gamma_{\rm sp} = 2, \, 2.25$ and $2.5$), for $m_\chi = 10$~GeV (blue curves) and $10^5$~GeV (red curves). In the presence of a DM spike, the $J$-factor asymptotically reaches a large value at a smaller value of $\theta_{\max}$, when compared against the NFW profile. }
    \label{fig:dm_jfactor}
\end{figure}

In practice, our results are controlled by how the ROI-integrated $J$-factor varies with the angle $\Delta\Omega$, the slope of the spike $\gamma_{\rm sp}$, DM mass $m_{\chi}$, and the annihilation cross-section $\langle \sigma v \rangle$.
In Fig.~\ref{fig:dm_jfactor}, we show the dependence of the $J$-factor on maximum angular cutoff $\theta_{\max}$, for a fixed $\langle \sigma v \rangle = 2.2 \times 10^{-26}~{\rm cm}^3~{\rm s}^{-1}$ corresponding to the canonical thermal relic freeze-out benchmark~\cite{Steigman:2012nb}.
In the presence of a GS spike, the $J$-factor asymptotically reaches a constant value at a much smaller $\theta_{\max}$, with the behavior even more prominent for larger DM masses. For example, for $m_\chi =10^5$~GeV, a $\theta_{\max} \approx 10^{-2}$~arcsec would be sufficient to capture the dominant DM annihilation contribution in the presence of a spike.
This is because for higher values of DM mass, $\rho_{\rm sat}$ is even larger, leading to a smaller saturation radius $R_{\rm sat}$, which enhances the $J$-factor further. If a DM annihilation signal were observed, its dependence on the ROI angular size would provide a direct probe of whether the underlying GC DM distribution is spiked or smooth. The current angular resolution of \emph{Fermi}-LAT is $<0.15^\circ$ at $>10$ GeV~\cite{fermi}, as shown by the rightmost vertical line in Fig.~\ref{fig:dm_jfactor}. Future gamma-ray telescopes such as APT~\cite{Buckley:2021LE} are expected to achieve 3--10 times better angular resolution, thus offering enhanced sensitivity to the spike profile, and in particular, a better handle on distinguishing between the spike and NFW (or cored) profiles.

In what follows, we consider the DM annihilation into photon channel and derive some of the strongest constraints to-date on the $\gamma_{\rm sp}$ parameter value in the WIMP thermal relic freeze-out scenario.

\section{Gamma-ray line searches and constraints on the spike}
\label{sec:gamma_lines}

A distinctive indirect signature of annihilating non-relativistic WIMPs is a monochromatic gamma-ray line,  arising from the process $\chi\chi\to\gamma\gamma$, which produces photons at $E_\gamma\simeq m_\chi$.
In Eq.~\eqref{eq:flux_master} this corresponds to a line-like spectrum $dN_\gamma/dE_\gamma\propto \delta(E_\gamma-m_\chi)$,
so that the energy dependence is fixed,  while the overall normalization is controlled by the product
$\langle\sigma v\rangle_{\gamma\gamma}\,J(\Delta\Omega)$.
In this section, we first recast the constraints obtained from \emph{Fermi}-LAT~\cite{Foster:2022nva} and MAGIC~\cite{MAGIC:2022acl} observations of the GC.

\subsection{\emph{Fermi}-LAT constraints}
\label{subsec:fermi_lines}

The \emph{Fermi} Large Area Telescope (LAT) is a space-based gamma-ray telescope that surveys the sky in the GeV range and above.
In Ref.~\cite{Foster:2022nva}, a dedicated search for DM annihilation into monochromatic $\gamma$-ray lines was performed using observation of the inner Galaxy with 14 years of \emph{Fermi}-LAT data between $10~\mathrm{GeV}$ and $2~\mathrm{TeV}$. This search uses a template-based likelihood analysis, with the analysis region taken up to $30^\circ$ away from the GC. 
For the search, a Galactic-plane mask is applied to reduce contamination from diffuse Galactic emission (see Fig. 1 left panel of Ref.~\cite{Foster:2022nva}). This removes the brightest plane emission at larger angular separations while leaving the central region unmasked.
Note that this $30^\circ$ region is not treated as a single integrated bin.
Instead, the region is divided into concentric $1^\circ$-wide annuli, and the analysis uses a joint likelihood constructed as the product of the per-annulus likelihoods.
This annular construction is important because different annuli have different foreground and background contamination and different signal-to-background ratios.

When the constraint is obtained from a joint likelihood over many annuli, a simple overall rescaling with the $J$-factor will not be able to properly translate it to a bound for a spiky DM profile.
This is because a spiked DM profile would change the relative signal normalization across annuli, strongly enhancing the innermost annulus while leaving the outer annuli essentially unaffected.
Therefore, recasting the published $(0^\circ-30^\circ)$ limit with a single $J$-factor (using Eq.~\eqref{eq:recast_equation}) would not, in general, reproduce the exact limit that would follow from a full reanalysis with a spike template. On the other hand, for a result derived from the innermost annulus ($0^\circ$--$1^\circ$), the translation is robust because it involves a single, well-defined $J$-factor in Eq.~\eqref{eq:recast_equation}.

\begin{figure}[t]
    \centering
    \includegraphics[width=0.6\linewidth]{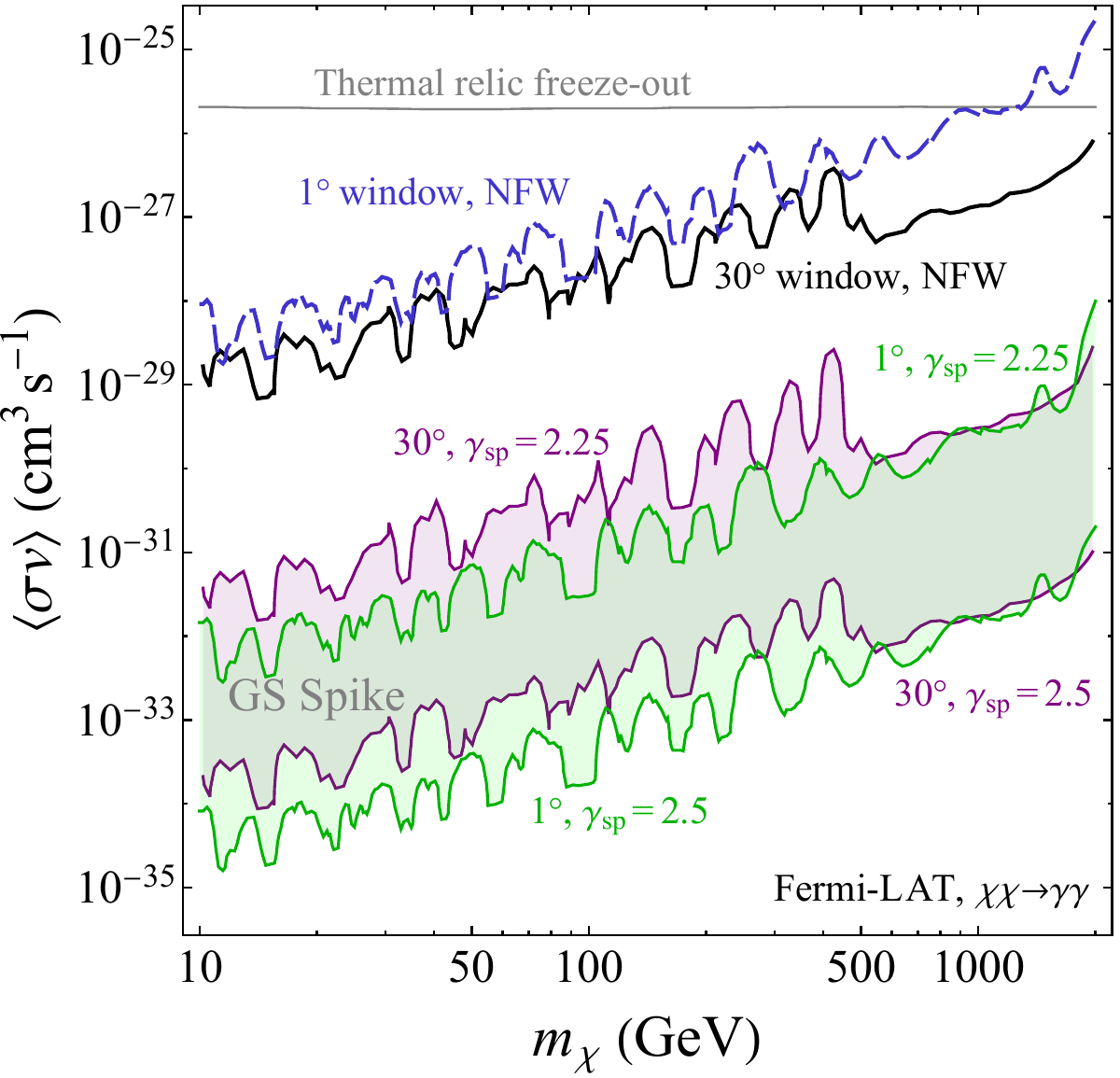}
    \caption{ Comparison of \emph{Fermi}-LAT constraints for two angular regions centered on the GC. The NFW constraints are taken from Ref.~\cite{Foster:2022nva}, where the blue (black) curve corresponds to constraints from observation within $1^\circ$ ($30^\circ$) of the GC.
    The green (purple) band corresponds to a recast of the NFW constraints to those corresponding to a spiky DM profile.
    In the presence of a steep DM spike, the relative enhancement of $J$-factor is larger at smaller $\theta_{\max}$. As a result, for steep spikes with $\gamma_{\rm sp}\in [2.25,\,2.5]$, the $(0-1)^\circ$ limits become stronger than those from $(0-30)^\circ$.}
    \label{fig:fermi_lat_comparison}
\end{figure}

Furthermore, a larger angular cut-off would mean that the enhancement in the $J$-factor due to the spike will be less pronounced, since at larger angular cut-offs, the ratio of the $J$-factors corresponding to a spiked and a standard NFW profile become smaller (see Fig.~\ref{fig:dm_jfactor}). In Fig.~\ref{fig:fermi_lat_comparison}, we compare the \emph{Fermi}-LAT constraints for two angular regions, $(0^\circ$--$30^\circ)$ and $(0^\circ$--$1^\circ)$, around the GC.
The blue curve corresponds to constraints from observation within $(0$–$1)^\circ$ of the GC (inner GC regions), while the black curve shows the constraint from the larger $(0$–$30)^\circ$ region, for the NFW profile~\cite{Foster:2022nva}.
We show the constraints in the presence of the spike (with $\gamma_{\rm sp}\simeq 2.25$–$2.5$, corresponding to a GS spike) using the green and purple curves, for the $(0-1)^\circ$ and $(0-30)^\circ$ regions, respectively. The semi-opaque shaded regions denote the results for intermediate values of $\gamma_{\rm sp}$.
For further discussions of how a DM spike impacts the \emph{Fermi}-LAT limits, including a comparison between the $(0^\circ$--$30^\circ)$ and $(0^\circ$--$1^\circ)$ regions, see Appendix~\ref{sec:appendixA}.

In the absence of strong central enhancement due to DM spike, the wider window yields stronger limits due to the larger signal statistics, so for an NFW profile, the results corresponding to the $(0$–$30)^\circ$ region (black curve) is stronger.
However, a GC-centered analysis is more sensitive to a DM spike.
As a result, for steep spikes with $\gamma_{\rm sp}\simeq 2.25$–$2.5$, the $(0-1)^\circ$ limits are observed to be stronger than those from $(0-30)^\circ$. 
For these two reasons stated above, we adopt the published inner $(0^\circ$--$1^\circ)$ \emph{Fermi}-LAT annulus result of Ref.~\cite{Foster:2022nva} as the baseline input for our spike analysis.

\subsection{MAGIC constraints}
\label{subsec:magic_lines}

MAGIC (Major Atmospheric Gamma Imaging Cherenkov Telescopes\footnote{Renamed to MAGIC Florian Goebel Telescopes shortly after the 2008 launch~\cite{2009arXiv0907.1211C}.}) is a system of 2 imaging atmospheric Cherenkov telescopes, situated on La Palma, $2.2$~km above the sea level. The $\gamma$-ray line search toward the GC by the MAGIC collaboration~\cite{MAGIC:2022acl} probes DM masses well above the \emph{Fermi}-LAT reach, extending up to DM mass of $m_\chi \sim \mathcal{O}(100)\,\mathrm{TeV}$.
It combines multiple observational configurations, which are associated with slightly different circular signal regions around the GC
(quoted radii include $0.5^\circ$, $1.0^\circ$, and $1.1^\circ$ in the analysis setup). The final MAGIC constraint corresponds to a maximum observation window (ROI) of $1.1^\circ$ centered around the GC. For steep spikes of interest, the enhancement of $J(\Delta\Omega)$ is largest at small $\theta_{\max}$ and changes only mildly once $\theta_{\max}$ is increased beyond the angular scale that encloses the spike emission (see Fig.~\ref{fig:dm_jfactor}). Therefore, to obtain a conservative estimate of the constraints in the presence of a spike, we rescale the MAGIC NFW line limits using Eq.~\eqref{eq:recast_equation}, for a maximum observation window of $1.1^\circ$ around the GC.

Note that the NFW reference profiles used by Ref.~\cite{Foster:2022nva} and Ref.~\cite{MAGIC:2022acl} employ slightly different numerical parameters.
For our analysis, we use the same NFW parameters as  in Ref.~\cite{Foster:2022nva} [cf.~Eq.~\eqref{eq:nfw_standard}], 
corresponding to a local DM density of $\rho_\odot = 0.4$~GeV cm$^{-3}$ from Eq.~\eqref{eq:NFW_halo}. On the other hand, in Ref.~\cite{MAGIC:2022acl}, DM annihilation constraints are obtained for an NFW profile with
\begin{equation}
      r_s^{\rm \tiny(MAGIC)}=21~{\rm kpc}\; , \quad R_\odot^{\rm \tiny(MAGIC)} = 8.5~{\rm kpc}\; , 
      \quad \rho_s^{\rm \tiny(MAGIC)} \approx 0.307~{\rm GeV\, cm}^{-3} \; ,
\end{equation}
corresponding to a local DM density of $\rho_\odot = 0.384$~GeV cm$^{-3}$ from Eq.~\eqref{eq:NFW_halo}, where a maximum halo size of $402$~kpc is considered.
Comparing the corresponding $J$-factors, we find that rescaling the limits of Ref.~\cite{MAGIC:2022acl} to our NFW parametrization in Eq.~\eqref{eq:nfw_standard} yields a factor of $0.88$ on the limits on $\langle \sigma v\rangle$, i.e.\ the MAGIC constraints become $12\%$ stronger.
This is because, even in the absence of any DM spike, our NFW parametrization leads to a $J$-factor which is $\sim 12\%$ larger.

Recast into a consistent NFW parametrization, in what follows, we show the effects of a DM spike, and calculate the maximum allowed value of the spike slope $\gamma_{\rm sp}$ as a function of DM mass, in the thermal relic freeze-out paradigm.

\subsection{Constraining DM annihilation and spike profile}
\label{subsec:gamma_sp_bounds}

\begin{figure}[t]
    \centering
    \includegraphics[height=0.45\linewidth]{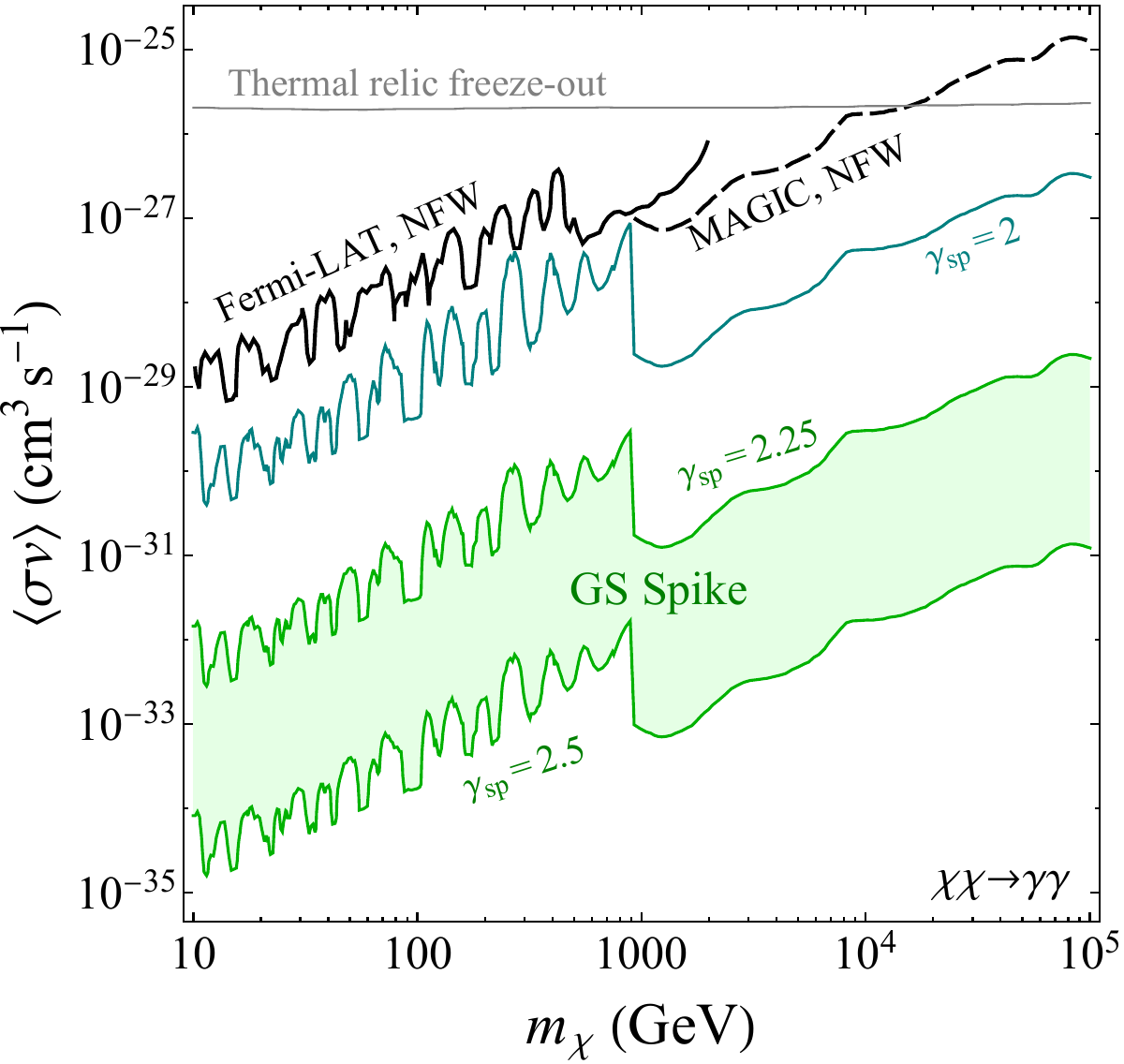}
    \hspace{0.5cm}
    \includegraphics[height=0.452\linewidth]{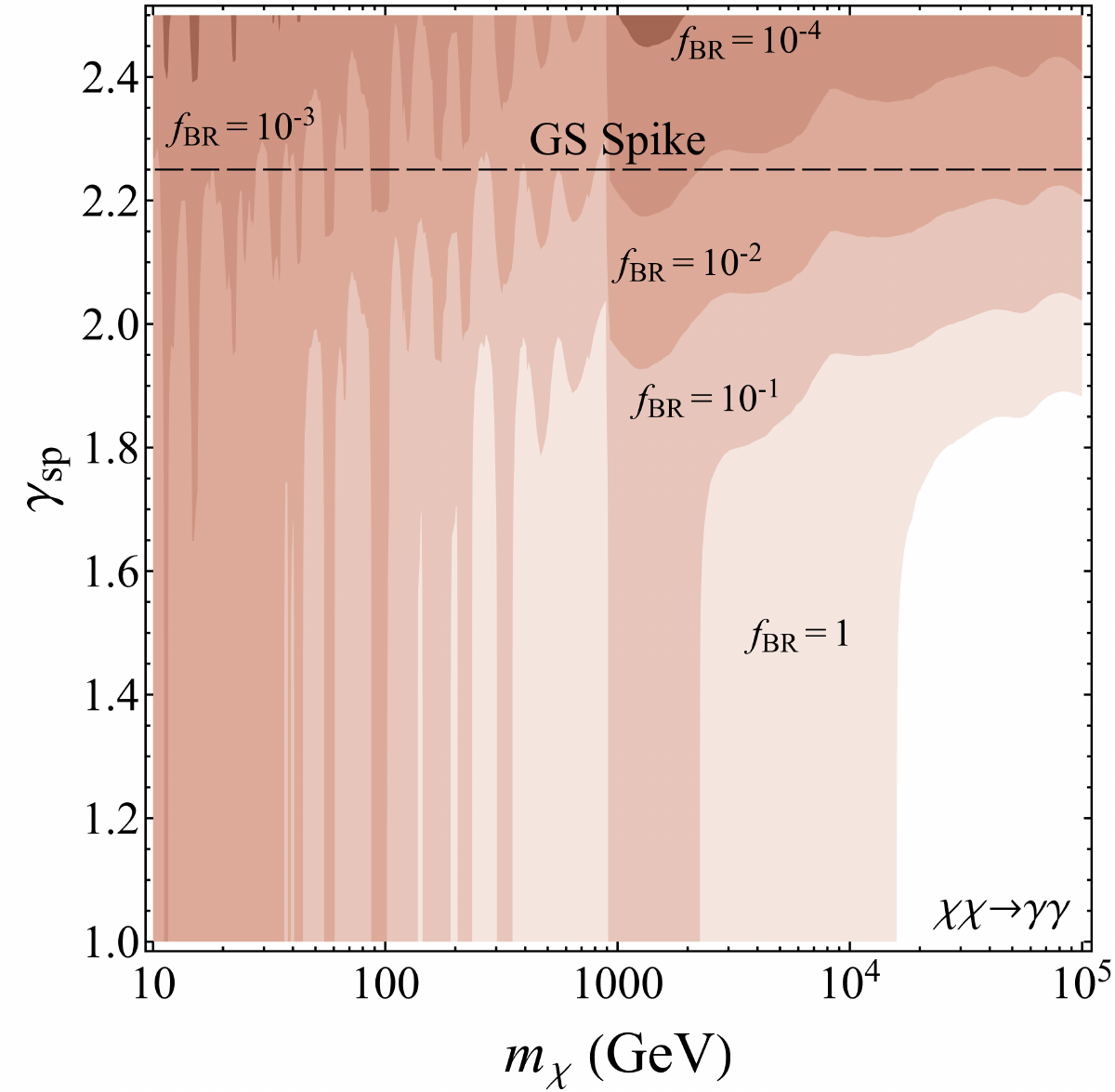}
    \caption{\textit{Left panel:} Constraints on DM annihilation cross section $\langle \sigma v\rangle$ from gamma-ray line searches toward the GC, as a function of the DM mass.
    The black solid line for \emph{Fermi}-LAT corresponds to an observation window of $(0-30)^\circ$ and the black dashed curve for MAGIC up to  $1.1^\circ$ around the GC, for an NFW profile.
    The cyan and green curves show our spike recast using Eq.~\eqref{eq:recast_equation} for GS-like spike benchmarks.
    For the rescaled \emph{Fermi}-LAT constraints, we obtain the results from the inner $(0-1)^\circ$ annulus results in Ref.~\cite{Foster:2022nva}. 
    \textit{Right panel:} Upper bounds on the spike slope  $\gamma_{\rm sp}$ for a thermal relic freeze-out WIMP, for DM annihilation to photons.
    The different shades correspond to photon-line annihilation branching ratios $f_{\rm BR}=10^{-n}$ with integer $n=[0,4]$ (lighter to darker shade). For $n\leq 2$, a GS spike with $\gamma_{\rm sp}\in [2.25, 2.5]$ is effectively ruled out for the thermal freeze-out scenario.
    }
    \label{fig:photon_bounds}
\end{figure}

Our combined line constraints from \emph{Fermi}-LAT and MAGIC, together with the spike reinterpretation, are summarized in Fig.~\ref{fig:photon_bounds} left panel.
The black solid and dashed curves show the NFW halo limits on $\langle \sigma v \rangle$.
The dark cyan curve corresponds to $\gamma_{\rm sp} =2$ constraints. The green curves denote the minimum ($\gamma_{\rm sp} = 2.25$) and maximum ($\gamma_{\rm sp}=2.5$) value for a GS spike, with the shaded light-green region denoting intermediate values.
For the \emph{Fermi}-LAT constraints in the presence of a spike  we use the inner $(0$--$1)^\circ$ result, as explained in Sec.~\ref{subsec:fermi_lines} and Fig.~\ref{fig:fermi_lat_comparison}.
For MAGIC we apply the same $J$-factor rescaling to the NFW limits,  corresponding to an observation window of $1.1^\circ$ around the GC, which provides the extension up to 100~TeV DM mass regime.\footnote{A thermal WIMP beyond ${\cal O}(100)$ TeV is anyway disfavored by the unitarity bound~\cite{Griest:1989wd}.} Note that, in the presence of a spike, the constraints can strengthen by up to a factor of $\mathcal{O}(10^6)$.

In the right panel of Fig.~\ref{fig:photon_bounds}, we show the constraint on the spike parameter $\gamma_{\rm sp}$, for a thermal freeze-out WIMP DM\footnote{We remind the reader, that for masses above $10$~GeV, the thermal freeze-out WIMP DM can account for all of the DM for an annihilation cross section of $\langle \sigma v \rangle \approx 2.2 \times 10^{-26}\,\mathrm{cm^3\,s^{-1}}$~\cite{Steigman:2012nb}.}.
The $\gamma$-ray line limits can be translated into constraints on the spike parameters by combining Eq.~\eqref{eq:recast_equation} with the steep dependence of $J_\chi(\Delta\Omega)$ on $\gamma_{\rm sp}$.
We present the results for 5 different values of the photon-line branching ratio (BR), defined as
\begin{equation}
    f_{\rm BR} \equiv \langle\sigma v\rangle_{\gamma\gamma}/ \langle\sigma v\rangle_{\rm tot} \; .
\end{equation}   
Here, $\langle\sigma v\rangle_{\gamma\gamma}$ denotes the annihilation cross section for the $\chi\chi \to \gamma\gamma$ process, and $\langle\sigma v\rangle_{\rm tot}$ denotes the cross section for $\chi\chi \to$~all, which we take as the thermal cross section $\langle \sigma v\rangle_{\rm th}$.
Such a scenario can be easily realized in the presence of a multicomponent dark sector, or even for scenarios where the DM dominantly annihilates into a far less ``visible'' sector, such as neutrinos. For $f_{\rm BR}<1$, the total annihilation cross section is given by $\langle\sigma v\rangle (\chi\chi \to \gamma\gamma)/f_{\rm BR}$. 
To translate an upper limit on the annihilation cross-section into an upper limit on the value of $\gamma_{\rm sp}$, we impose the condition that even in the presence of a spike, the DM annihilation cross-section must be fixed at the thermal relic freeze-out value. Therefore, for a certain value of the spike parameter $\gamma_{\rm sp}$, if the condition 
\begin{equation}
    J_{\rm spike} \langle \sigma v \rangle_{\rm th} \leq J_{\rm NFW} \langle \sigma v \rangle_{\rm NFW}
\end{equation}
is satisfied, then that value of $\gamma_{\rm sp}$ would be allowed. Here, $\langle \sigma v \rangle_{\rm NFW}$ denotes the existing constraints for an NFW profile, $\langle \sigma v \rangle_{\rm th}$ is the thermal relic annihilation cross-section for the freeze-out scenario, and $J_{\rm spike}$ is a function of $m_\chi$, $\gamma_{\rm sp}$, and the angular observation window $\Delta \Omega$.
Since the $J$-factor grows rapidly with $\gamma_{\rm sp}$ through the scaling discussed in section~\ref{subsec:Jfactor} and Eq.~\eqref{eq:jchi_scaling},  larger $\gamma_{\rm sp}$ values lead to stronger constraints on the quantity $\langle\sigma v\rangle_{\rm tot}=\langle\sigma v\rangle/f_{\rm BR}$, which directly map  into an upper limit on $\gamma_{\rm sp}$. Larger $f_{\rm BR}$ allows a stronger constraint on the total annihilation cross section,  and therefore, even for smaller values of $\gamma_{\rm sp}$, the constraint on the annihilation cross section becomes smaller than the thermal freeze-out value, excluding progressively smaller spikes as $f_{\rm BR}$ increases to 1.
Even for $f_{\rm BR} =10^{-2}$, a GS-like spike is almost completely excluded, with \emph{Fermi}-LAT constraints contributing at lower masses and MAGIC providing the leading sensitivity in the $\mathcal{O}(1-100)$~TeV regime.
It is remarkable that even with  $f_{\rm BR}$ as small as $10^{-4}$, we still constrain some portions of the spike parameter space.  

To put it in perspective, one may naively expect that the DM annihilation to photons is loop-suppressed with $f_{\rm BR}\sim (\alpha/4\pi)^2\sim 10^{-7}$, where $\alpha$ is the fine-structure constant. However, a full one-loop calculation of the neutralino DM annihilation to two photons in the minimal supersymmetric SM predicts $f_{\rm BR}$ typically between $10^{-4}$ to $10^{-2}$~\cite{Bergstrom:1997fh, Bern:1997ng}, and even higher $f_{\rm BR}\sim 10^{-1}$ is possible for bino DM in low-scale superstring models~\cite{Anchordoqui:2009bn}. It is also interesting to note that the DM annihilation interpretation of the Fermi-LAT gamma-ray line excess requires a best-fit value of $f_{\rm BR}\gtrsim 10^{-1}$~\cite{Tempel:2012ey,Weniger:2012tx}, and various WIMP DM scenarios have been considered that can accommodate such a large $f_{\rm BR}$; see e.g. Refs.~\cite{Bertone:2009cb,Cline:2012nw,Choi:2012ap,Lee:2012bq,Buckley:2012ws,Bai:2012qy}. According to Fig.~\ref{fig:photon_bounds}, the possibility of a GC spike would be ruled out for all these WIMP DM models.

Having discussed DM annihilation to photons, in the next section, we explore the neutrino channel, which is experimentally far more challenging to probe.
But it comes with the advantage that the ROI is very large and almost always includes the GC in such analyses, thus making the neutrinos a complementary probe of the GC spike.

\section{Neutrino line searches and constraints on the spike}
\label{sec:nu_lines}

A complementary probe of GC annihilation is provided by neutrino telescopes, which can search for the direct channel
$\chi\chi\to\nu\bar\nu$. This produces a neutrino line at $E_\nu\simeq m_\chi$ (broadened by detector resolution).
Here we use the IceCube GC line search of Ref.~\cite{IceCube:2023ies}, which reports upper limits on
$\langle\sigma v\rangle_{\nu_e\bar\nu_e}$ over a wide mass range assuming an NFW halo profile\footnote{The corresponding limits for $\nu_\mu\bar\nu_\mu$ and $\nu_\tau\bar\nu_\tau$ are somewhat weaker. Therefore, we present results only for the $\nu_e\bar{\nu}_e$ channel. Further, we do not include the latest IceCube limits based on 9 years of IceCube-DeepCore data~\cite{IceCube:2025fcn}, since these are restricted to the $5-200$ GeV DM mass range~\cite{IceCube:2025fcn}, whereas the previous analysis~\cite{IceCube:2023ies} adopted here covers a wider DM mass range.}.
Although Ref.~\cite{IceCube:2023ies} does not quote the NFW halo parameters used, we assume that these are inherited from a previous IceCube analysis~\cite{IceCube:2015rnn} where the mean values used are as follows:
\begin{equation}
    r_s = 16.1~{\rm kpc}\; , 
    \quad R_{\odot} = 8.5~{\rm kpc}\; ,
    \quad \rho_s = 0.532~{\rm GeV\, cm}^{-3}\; ,
\end{equation}
corresponding to a local DM density of $\rho_{\odot} =0.471~{\rm GeV \, cm}^{-3}$. We consider a maximum halo size of $400$~kpc. The DM density at $400$~kpc is approximately $10^4$ times smaller than the local DM density, at $3.9\times 10^{-5}~{\rm GeV\,cm^{-3}}$; Since the annihilation signal scales with the square of the DM density, our results are not sensitive to the exact choice of the DM halo size.
Rescaling the limits of our NFW parameterization [see Eq.~\eqref{eq:nfw_standard}], we obtain a $36\%$ weaker constraint compared to that presented in~Ref.~\cite{IceCube:2023ies}.

To constrain DM annihilation into neutrinos from IceCube data, Ref.~\cite{IceCube:2023ies} performs a likelihood analysis that uses how the signal and background populate different regions of the reconstructed energy ($E_{\rm rec}$) and angular separation $(\theta_{\rm rec})$ from the GC.
A strict reinterpretation in the presence of a spiky DM profile would therefore require repeating the IceCube analysis with the spike morphology inserted directly into the signal template.
In particular, a spike template would make the signal significantly more concentrated towards the GC than the NFW template used in Ref.~\cite{IceCube:2023ies}, and a dedicated analysis could, in principle, capitalize on this additional spatial contrast to improve discrimination.
Our approach instead keeps the published IceCube analysis fixed and modifies only the overall signal normalization through the corresponding $J$-factor ratio.  This makes our approximate reinterpretation typically more conservative by construction.

For our conservative approximation, we solve Eq.~\eqref{eq:recast_equation} with $J$-factors computed in a ROI defined by $\theta_{\max}=50^\circ$~\footnote{Fig.~3 of Ref.~\cite{IceCube:2023ies} (shown for $m_\chi=1~\mathrm{TeV}$) indicates that the reconstructed signal probability density for cascade-like events is concentrated within $\theta_{\rm rec}\lesssim 50^\circ$. This choice is conservative for the higher DM mass ranges that are most relevant for the IceCube constraints, since the angular reconstruction would improve further with higher neutrino energy.}.
We also assume a $100\%$ branching fraction into neutrinos, so that the IceCube bound can be interpreted directly as a limit on $\langle\sigma v\rangle$ in the presence of a DM spike in the GC.

\begin{figure}[t!]
    \centering
    \includegraphics[height=0.45\linewidth]{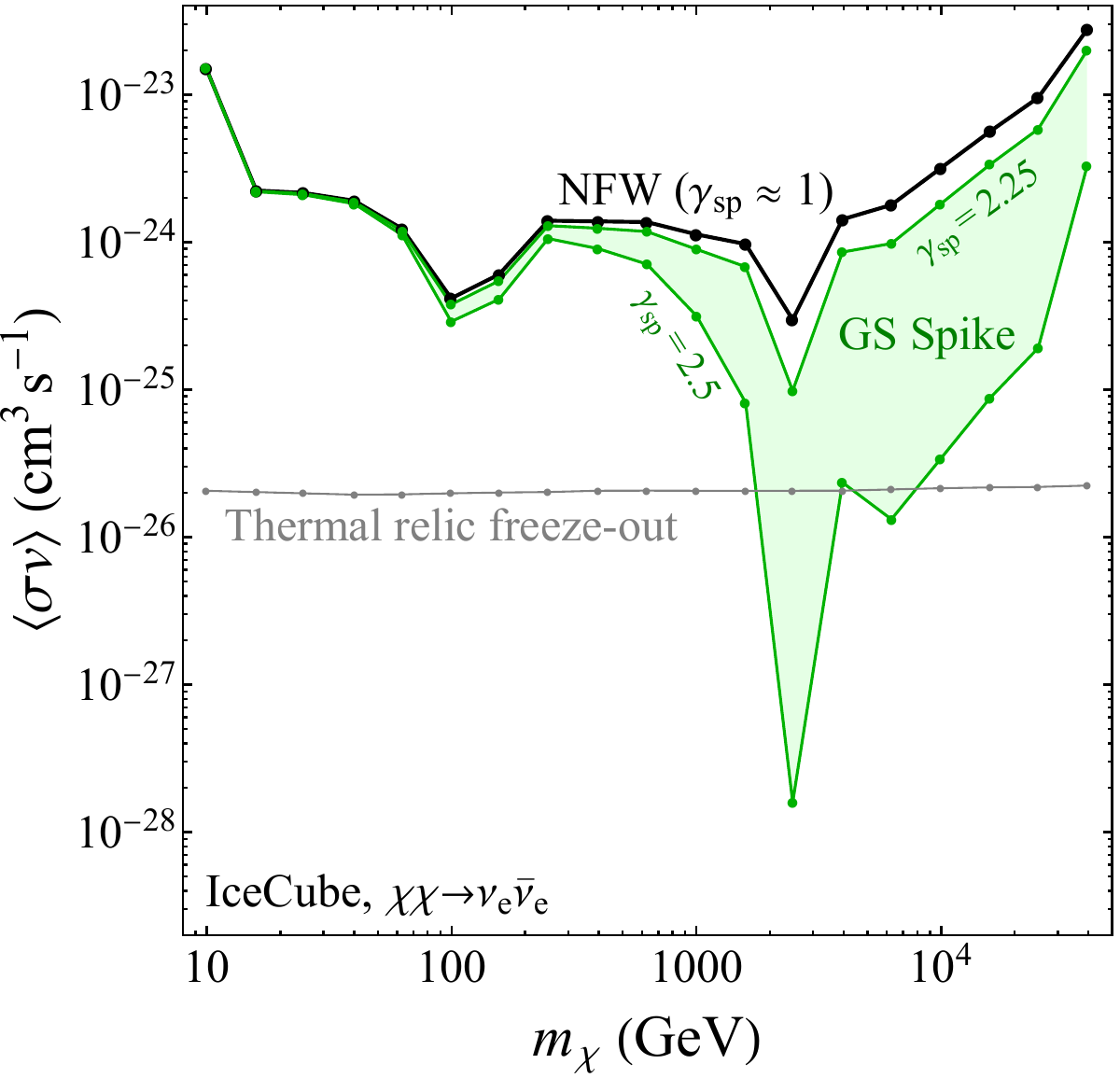}
    \hspace{0.5cm}
    \includegraphics[height=0.455\linewidth]{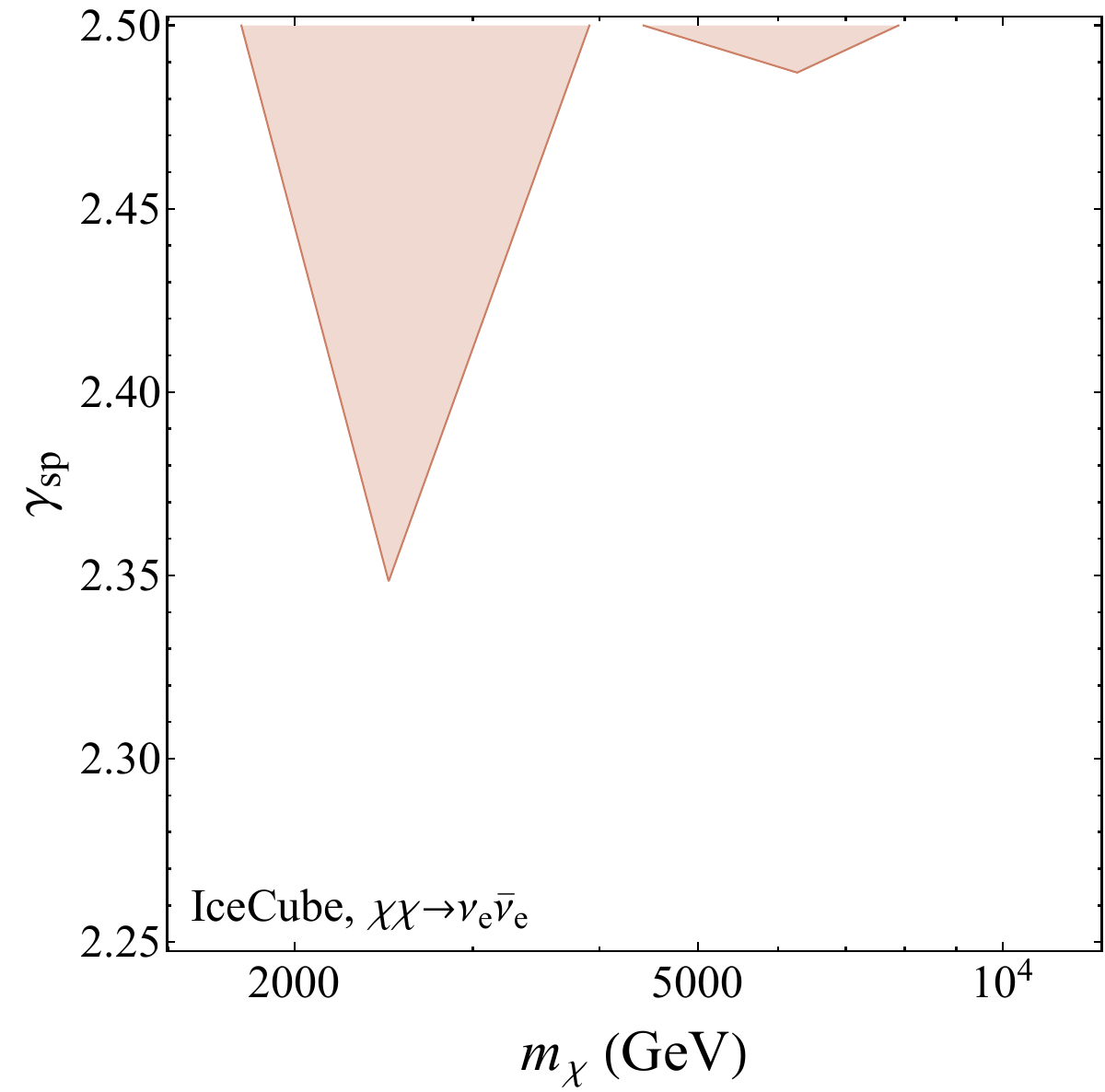}
    \caption{\textit{Left panel:} Constraints on DM annihilation to neutrinos from IceCube.
    The black curve shows the IceCube limit obtained for the NFW profile parametrization used in this work, while the green curves show our spike reinterpretation for GS-like spike slopes of $\gamma_{\rm sp}=2.25$ and $\gamma_{\rm sp} = 2.5$, with the shaded region denoting intermediate values of $\gamma_{\rm sp}$. The thermal relic benchmark is shown in gray.
    \textit{Right panel:} Constraints on $\gamma_{\rm sp}$ assuming a thermal relic  freeze-out annihilation cross section into neutrinos, shown as a function of the DM mass $m_\chi$.
    The strongest constraint occurs in the $m_\chi \sim (1.8- 3.9)$~TeV window, reaching $\gamma_{\rm sp}\simeq 2.35$.}
    \label{fig:neutrino_bounds}
\end{figure}

Our results are summarized in Fig.~\ref{fig:neutrino_bounds}.
The left panel shows the IceCube NFW limits (black) together with the corresponding limits in the presence of a DM spike (green) for GS-like slopes ($\gamma_{\rm sp}=2.25$, and $\gamma_{\rm sp}=2.5$), illustrating that the presence of a spike can substantially strengthen the DM annihilation constraints, in particular in the $\mathcal{O}(1-50)$~TeV mass range. For a detailed discussion on how the shape of the IceCube constraint changes in the presence of a DM spike, see Appendix~\ref{sec:appendixA}.
The right panel shows the implied constraints on the spike parameter $\gamma_{\rm sp}$ obtained by fixing the allowed $\langle\sigma v\rangle$ to the thermal relic freeze-out value, and solving, as a function of $m_\chi$, for the maximum $\gamma_{\rm sp}$ that still allows the thermal relic freeze-out value for the annihilation cross-section.
The strongest bound occurs in the few-TeV window, reaching $\gamma_{\rm sp}\simeq 2.35$ in our setup. As expected, this is still a much weaker constraint compared to the gamma-ray constraints shown in Fig.~\ref{fig:photon_bounds}.

\section{Conclusions}
\label{sec:conclusion}

The existence of a dark matter spike around our Galactic Center remains unsettled.
In this paper, we derived new constraints on the Gondolo-Silk spike parameter $\gamma_{\rm sp}$ using the existing gamma-ray and neutrino line searches from the GC region. To this effect, we addressed two questions:
\begin{enumerate}
    \item [$(i)$] How are the DM annihilation cross-section bounds modified in the presence of a central spike?
    \item [$(ii)$] What do the current constraints on the annihilation cross-section imply for the spike parameters?
\end{enumerate}
To answer question $(i)$, we recast the experimental constraints on the thermally-averaged annihilation rate $\langle \sigma v\rangle$ derived using an NFW profile to those corresponding to the spike profile. Our results, shown in the left panels of Figs.~\ref{fig:photon_bounds} and \ref{fig:neutrino_bounds} for the gamma-ray and neutrino channels respectively, reflect orders of magnitude improvement in the $\langle\sigma v\rangle$ constraint in presence of a spike profile, which directly follows from the fact that the annihilation $J$-factor is proportional to the square of the DM density. 

To answer question $(ii)$, we assume the standard thermal WIMP paradigm which requires $\langle \sigma v\rangle\approx 2.2\times 10^{-26} {\rm cm}^3\, {\rm s}^{-1}$, and  translate our derived $\langle\sigma v\rangle$ constraints into limits of the GS spike parameter as a function of the DM mass. This is shown on the right panels of Figs.~\ref{fig:photon_bounds} and \ref{fig:neutrino_bounds} for the gamma-ray and neutrino channels, respectively.
We find that in the gamma-ray annihilation channel, the entire range of GS spike parameter $\gamma_{\rm sp}\in[2.25,2.5]$ is strongly constrained for photon branching fractions down to $\sim 10^{-2}$. In fact, sensitivity to  GS spike persists even for branching fractions as small as $\sim 10^{-4}$. To the best of our knowledge, this is the first indirect detection constraint on the spike parameter $\gamma_{\rm sp}$ as a function of the DM mass.
The corresponding limit from the neutrino channel turns out to be much weaker, which is expected, because the original $\langle \sigma v\rangle$ limits in the neutrino channel are orders of magnitude weaker than those in the photon channel. 

We focused our analysis on the thermal WIMP scenario mainly because $(i)$ this is by far the most popular choice, and $(ii)$ it gives a concrete model-independent theoretical benchmark value of $\langle \sigma v\rangle$ to make comparisons with.
Similarly, we only considered the photon and neutrino channels, as these typically provide the strongest and weakest limits, respectively. Nevertheless, one can easily extend our analysis to other annihilation channels as well as other DM scenarios. For instance, in the freeze-in case~\cite{Hall:2009bx}, the corresponding theoretical benchmark values of $\langle \sigma v\rangle$ are roughly $\sim 20$ orders of magnitude smaller than the freeze-out case (depending on the initial conditions)~\cite{Dev:2013oiy}. This is almost like the non-annihilating scenario, as far as the profile of spike in Fig.~\ref{fig:dm_density} is concerned. Depending on the model construction, observable indirect detection signals can still be expected from the GC region with high DM density~\cite{Heikinheimo:2018duk, Cosme:2020mck}. A detailed exploration of the implications of these beyond-WIMP scenarios on the existence of a spike profile is left for future work.

\acknowledgments
We wish to acknowledge the Center for Theoretical Underground Physics and Related
Areas (CETUP*) and the Institute for Underground Science at SURF for hospitality and for providing a stimulating environment during the 2025 Summer Workshop, where this work was initiated. We thank Jim Buckley, Chris Cappiello, Francesc Ferrer, Gonzalo Herrera, Matheus Hostert, Alejandro Ibarra, and Stephan Meighen-Berger for useful discussions. 
The work of B.D. was partly supported by the US Department of Energy under grant No. DE-SC0017987 and by a Humboldt Fellowship from the Alexander von Humboldt Foundation.
Y.P. acknowledges the support by the DFG Collaborative Research Institution Neutrinos and Dark Matter in Astro- and Particle Physics (SFB 1258).


\appendix

\section{Dependence of the $J$-factor on the DM mass and cross-section}
\label{sec:appendixA}

In this section, we illustrate how the presence of a DM spike can significantly strengthen annihilation constraints. To this end, we calculate the ratio of the annihilation $J$-factors in the spiked and NFW scenarios, for both the photon and neutrino channels.

\begin{figure}[t!]
        \centering
        \includegraphics[width=0.48\linewidth]{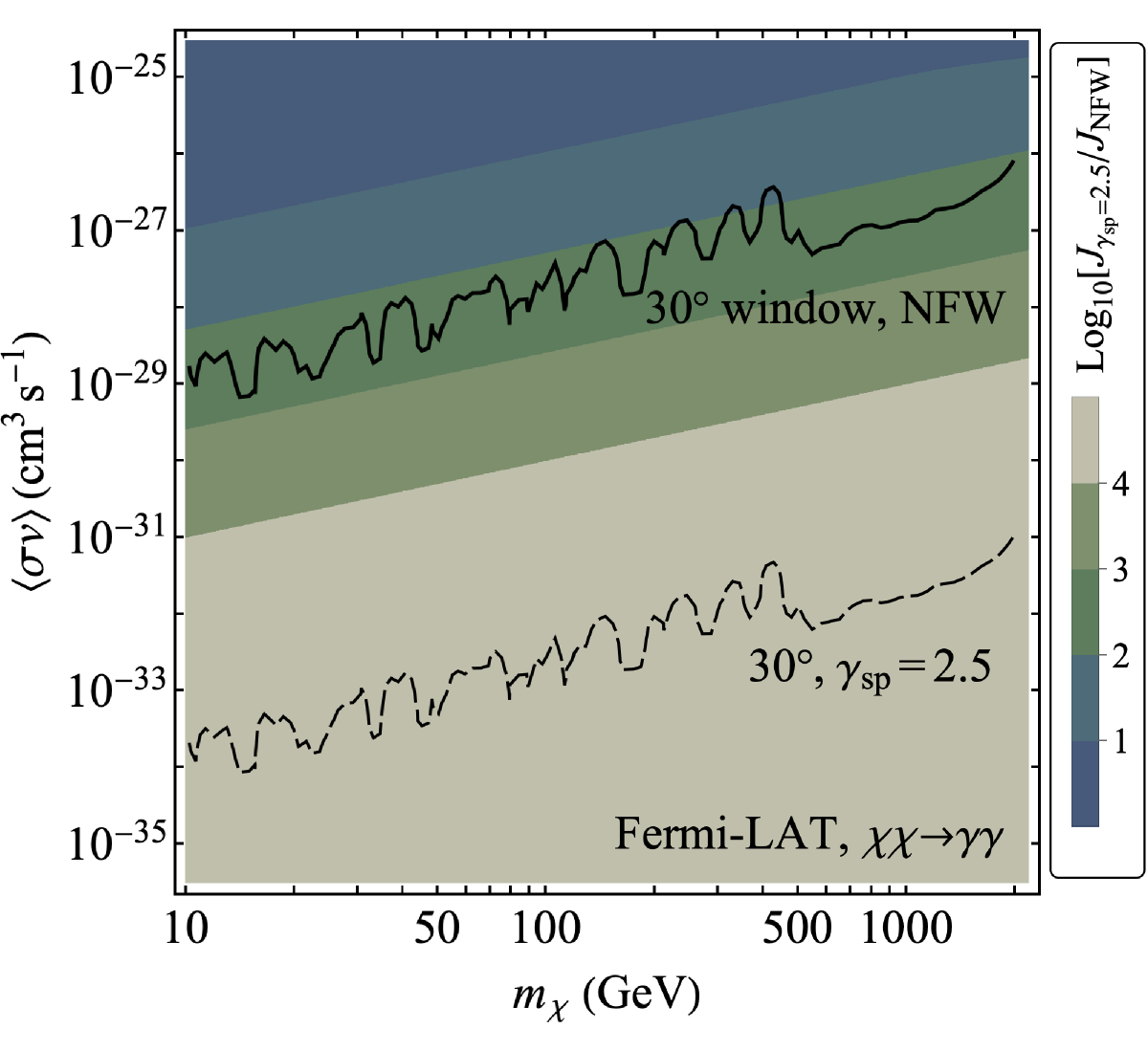}
        \hspace{0.25cm}
        \includegraphics[width=0.48\linewidth]{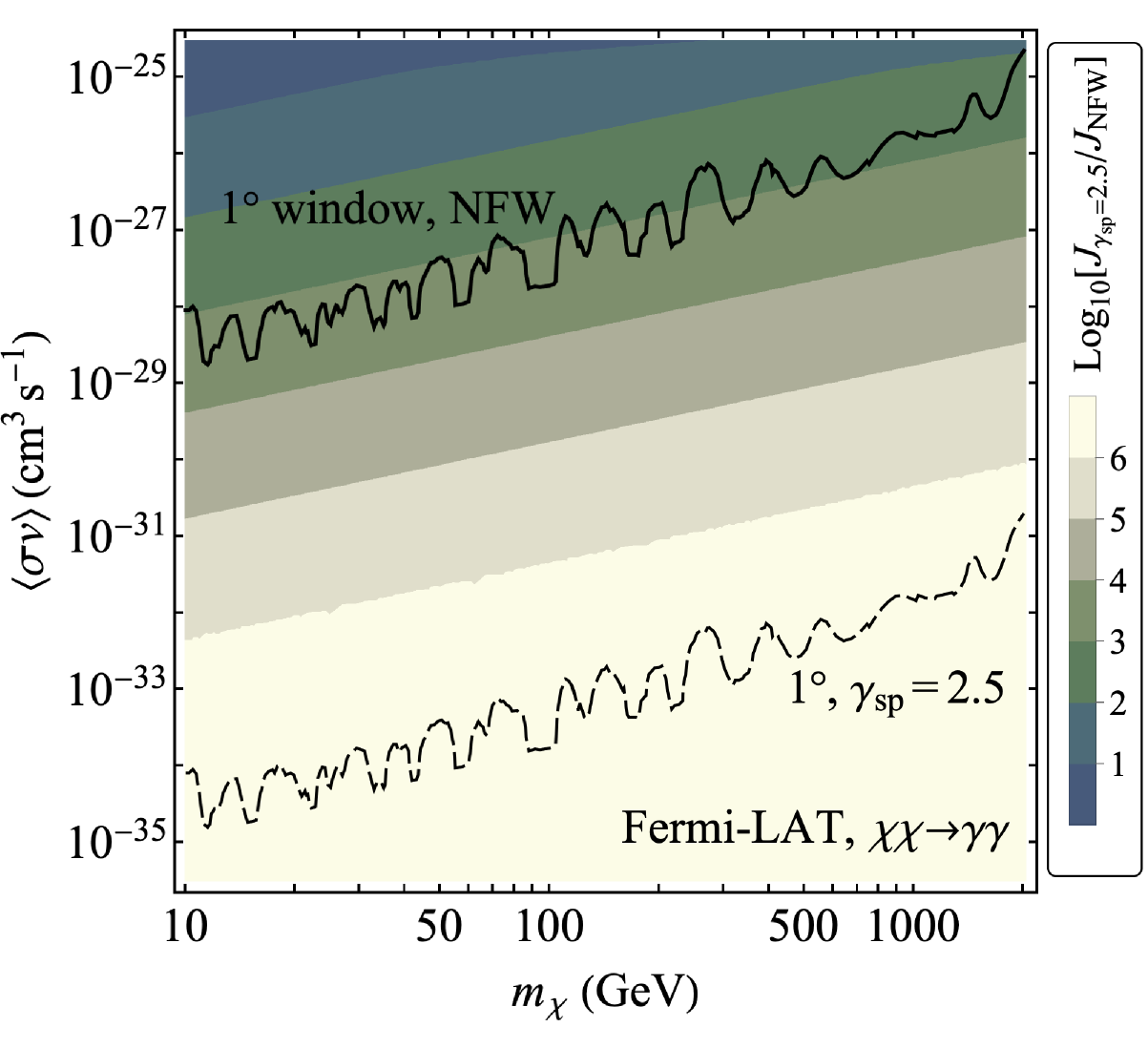}
        \caption{The ratio of the annihilation $J$-factors between a spiked profile (with $\gamma_{\rm sp}=2.5$) and an NFW profile is shown by the heat map. The black solid (dashed) line corresponds to the \emph{Fermi}-LAT constraints on $\chi\chi\to \gamma\gamma$ in the absence (presence) of a DM spike. The left (right) panel shows the results for a $30^\circ$ ($1^\circ$) observation window around the GC.}
        \label{fig:j_ratio_fermilat}
\end{figure}

In Fig.~\ref{fig:j_ratio_fermilat}, we show the results for \emph{Fermi}-LAT. In the left panel, we present the ratio of the $J$-factors as a function of DM mass and annihilation cross-section, for the $(0-30)^\circ$ region employed in Ref.~\cite{Foster:2022nva} (with the galactic plane away from the GC masked).
In the right panel, we show the same ratio for a $(0-1)^\circ$ region around the GC. By the black solid (dashed) curve, we show the constraints for DM annihilation to photons, for the NFW (spike, with $\gamma_{\rm sp} = 2.5$) profile.
The $J$-factor ratio reaches a maximum value of $\sim \mathcal{O}(10^4)$ for the $30^\circ$ region, whereas it reaches $\sim \mathcal{O}(10^6)$ for the $1^\circ$ region. This is expected, since in a smaller angular window, the spike contribution can be much larger relative to the NFW scenario.

In the presence of a DM spike, the enhanced $J$-factor shifts the NFW limit to even smaller values of $\langle \sigma v \rangle$.
Since the spike $J$-factor itself depends on $\langle\sigma v\rangle$ through the saturation density and saturation radius, moving to smaller $\langle\sigma v\rangle$ can further increase $J_{\rm spike}$ (see section~\ref{sec:spike-profile} for further discussions).
Therefore, the final constraint is reached once either the $J$-factor ratio has effectively saturated, or the relation $\langle \sigma v\rangle_{\rm NFW} J_{\rm NFW} = \langle \sigma v\rangle_{\rm spike} J_{\rm spike}$ is satisfied.
Since the $J$-factor ratio is larger by a factor of $\mathcal{O}(100)$ for the $1^\circ$ region compared to the $30^\circ$ region, the DM annihilation constraint is also strengthened more in the presence of a spike.
As a result, the $1^\circ$ limit can become stronger than the $30^\circ$ limit in the spiked case, even though the   $1^\circ$ NFW constraint is weaker.

\begin{figure}[t!]
        \centering
        \includegraphics[width=0.48\linewidth]{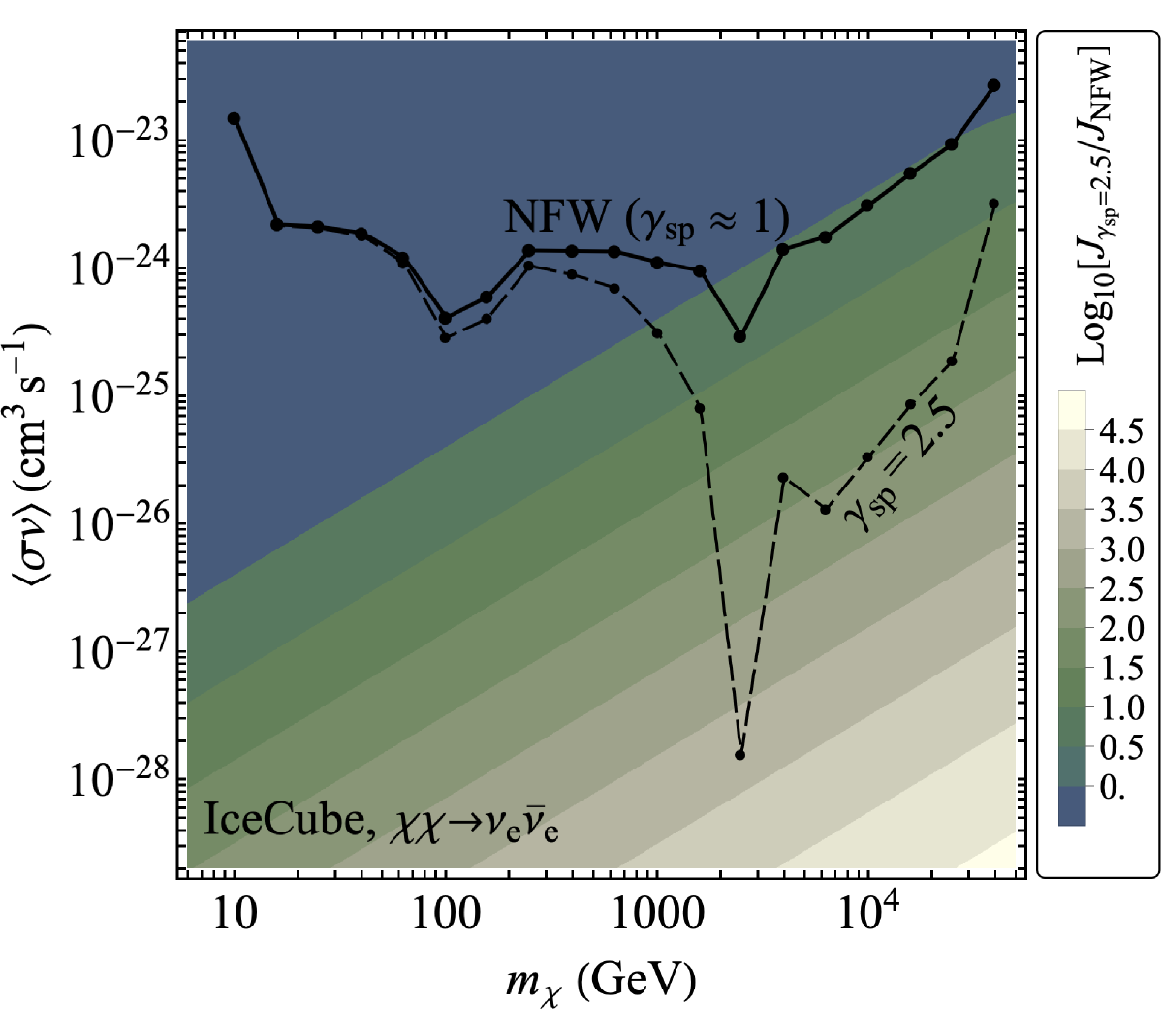}
        \caption{Ratio of the $J$-factors for a spiked DM profile (with $\gamma_{\rm sp}=2.5$) relative to an NFW profile, for an observation window of $50^\circ$ around the GC. The black solid (dashed) curve corresponds to the IceCube constraints on $\chi\chi\to \nu\bar{\nu}$ in the absence (presence) of a DM spike.}
        \label{fig:j_ratio_icecube}
\end{figure}

In Fig.~\ref{fig:j_ratio_icecube}, we show the $J$-factor ratio for IceCube, corresponding to a $50^\circ$ observation region around the GC.
We observe that the ratio can reach values as large as $\mathcal{O}(10^3)$ before the condition $\langle \sigma v\rangle_{\rm NFW} J_{\rm NFW} = \langle \sigma v\rangle_{\rm spike} J_{\rm spike}$ is satisfied, this leads to a substantial strengthening of the bound around $m_\chi\sim 2~\mathrm{TeV}$.
Note that the shape of the constraint changes visibly in the presence of a DM spike; this is due to the mass dependence of the $J$-factor.
Up to a mass of $\sim \mathcal{O}(100)$~GeV, and for $\langle \sigma v \rangle \sim (10^{-23}-10^{-24})~{\rm cm}^3~{\rm s}^{-1}$, the spike does not significantly enhance the $J$-factor. 
However, at higher masses, the same $\langle \sigma v \rangle$ can lead to a comparatively larger $J$-factor ratio, which shifts the limit to smaller $\langle \sigma v\rangle$ and thereby increases $J_{\rm spike}$ further.
As a result, the spike-induced strengthening is non-trivial and depends on the value of the $J$-factor ratio in the $(m_\chi, \langle \sigma v\rangle)$ plane.
This can be observed in Fig.~\ref{fig:j_ratio_icecube} in the $\sim 2$--$6$~TeV mass range, where, in the presence of a DM spike, the limits at $m_\chi \sim 2$~TeV and $m_\chi \sim 6$~TeV significantly strengthen when compared against the limit for $m_\chi \sim 4$~TeV, which shows a less pronounced enhancement. This leads to a dip in the constraint at $m_\chi \sim 4$~TeV, which is only present for the spiked scenario.

\bibliographystyle{JHEP}
\bibliography{main}

\end{document}